\documentclass[preprint,12pt]{elsarticle}

\usepackage{ifthen} 
\newboolean{pdflatex}
\setboolean{pdflatex}{true} 

\newboolean{articletitles}
\setboolean{articletitles}{true} 

\newboolean{uprightparticles}
\setboolean{uprightparticles}{false} 

\newboolean{inbibliography}
\setboolean{inbibliography}{false} 

\usepackage{booktabs}
\usepackage[utf8]{inputenc}
\usepackage{xspace}
\usepackage[separate-uncertainty = true]{siunitx}
\usepackage{hhline}
\usepackage{collcell}
\usepackage{tabularx}
\usepackage{multirow}
\usepackage{amssymb}
\usepackage{amsmath}
\usepackage{amsfonts}
\usepackage{comment}
\usepackage{lineno}
\usepackage[colorinlistoftodos]{todonotes}
\usepackage{makecell}
\usepackage{microtype}
\usepackage{caption} 
\usepackage{graphicx}
\usepackage{color}
\usepackage{colortbl}
\usepackage{pdflscape}
\usepackage{upgreek} 
\usepackage{natbib}
\usepackage{hyperref}
\usepackage{cleveref}
\usepackage{subcaption}

\graphicspath{{./figs/}} 
\usepackage[nobiblatex]{xurl}

\usepackage{xspace} 
\usepackage{upgreek}

\def\lhcb   {\mbox{LHCb}\xspace}

\def\cern {\mbox{CERN}\xspace}
\def\lhc    {\mbox{LHC}\xspace}

\def\upgradetwo {\mbox{Upgrade~II}\xspace}

\def\MagUp {\mbox{\em Mag\kern -0.05em Up}\xspace}

\ifthenelse{\boolean{uprightparticles}}%
{

 \def\PDelta      {\ensuremath{\Delta}\xspace}                 
 \def\PXi         {\ensuremath{\Xi}\xspace}                 
 \def\PLambda     {\ensuremath{\Lambda}\xspace}                 
 \def\PSigma      {\ensuremath{\Sigma}\xspace}                 
 \def\POmega      {\ensuremath{\Omega}\xspace}                 
 \def\PUpsilon    {\ensuremath{\Upsilon}\xspace}
 \let\oldPi\Pi
 \def\PPi         {\ensuremath{\oldPi}\xspace}

 \def\PB      {\ensuremath{\mathrm{B}}\xspace}                 
                  
 \def\PD      {\ensuremath{\mathrm{D}}\xspace}

 \def\PK      {\ensuremath{\mathrm{K}}\xspace}

 \def\Pi      {\ensuremath{\mathrm{i}}\xspace}

 \def\Ps      {\ensuremath{\mathrm{s}}\xspace}

 \def\thebaroffset{0.0em}
}
{

 \mathchardef\PDelta="7101
 \mathchardef\PXi="7104
 \mathchardef\PLambda="7103
 \mathchardef\PSigma="7106
 \mathchardef\POmega="710A
 \mathchardef\PUpsilon="7107
 \mathchardef\PPi="7105
                  
 \def\PB      {\ensuremath{B}\xspace}                 
                  
 \def\PD      {\ensuremath{D}\xspace}

 \def\PK      {\ensuremath{K}\xspace}

 \def\Pi      {\ensuremath{i}\xspace}

 \def\Ps      {\ensuremath{s}\xspace}

 \def\thebaroffset{0.18em}
}
\newcommand{\offsetoverline}[2][\thebaroffset]{\kern #1\overline{\kern -#1 #2}}%

\makeatletter
\ifcase \@ptsize \relax
  \newcommand{\miniscule}{\@setfontsize\miniscule{4}{5}}
\or
  \newcommand{\miniscule}{\@setfontsize\miniscule{5}{6}}
\or
  \newcommand{\miniscule}{\@setfontsize\miniscule{5}{6}}
\fi
\makeatother

\DeclareRobustCommand{\optbar}[1]{\shortstack{{\miniscule (\rule[.5ex]{1.25em}{.18mm})}
  \\ [-.7ex] $#1$}}

\def\squark    {{\ensuremath{\Ps}}\xspace}

\def\KorKbar {\kern \thebaroffset\optbar{\kern -\thebaroffset \PK}{}\xspace}

\def\D       {{\ensuremath{\PD}}\xspace}

\def\DorDbar {\kern \thebaroffset\optbar{\kern -\thebaroffset \PD}\xspace}

\def\Dp      {{\ensuremath{\D^+}}\xspace}
\def\Dm      {{\ensuremath{\D^-}}\xspace}

\def\DpDm    {\ensuremath{\Dp {\kern -0.16em \Dm}}\xspace}

\def\B       {{\ensuremath{\PB}}\xspace}

\def\BorBbar {\kern \thebaroffset\optbar{\kern -\thebaroffset \PB}\xspace}

\def\Bd      {{\ensuremath{\B^0}}\xspace}

\def\BdorBdbar {\kern \thebaroffset\optbar{\kern -\thebaroffset \Bd}\xspace}

\def\Bs      {{\ensuremath{\B^0_\squark}}\xspace}

\def\BsorBsbar {\kern \thebaroffset\optbar{\kern -\thebaroffset \Bs}\xspace}

\def\Y#1S{\ensuremath{\PUpsilon{(#1S)}}\xspace}

\def\LorLbar     {\kern \thebaroffset\optbar{\kern -\thebaroffset \PLambda}\xspace}

\def\AT#1     {\ensuremath{A_{\mathrm{T}}^{#1}}\xspace}           

\def\C#1      {\ensuremath{\mathcal{C}_{#1}}\xspace}                       
\def\Cp#1     {\ensuremath{\mathcal{C}_{#1}^{'}}\xspace}                    
\def\Ceff#1   {\ensuremath{\mathcal{C}_{#1}^{\mathrm{(eff)}}}\xspace}        
\def\Cpeff#1  {\ensuremath{\mathcal{C}_{#1}^{'\mathrm{(eff)}}}\xspace}       
\def\Ope#1    {\ensuremath{\mathcal{O}_{#1}}\xspace}                       
\def\Opep#1   {\ensuremath{\mathcal{O}_{#1}^{'}}\xspace}                    

\newcommand{\aunit}[1]{\ensuremath{\text{\,#1}}}       

\newcommand{\tev}{\aunit{Te\kern -0.1em V}\xspace}
\newcommand{\gev}{\aunit{Ge\kern -0.1em V}\xspace}
\newcommand{\mev}{\aunit{Me\kern -0.1em V}\xspace}
\newcommand{\kev}{\aunit{ke\kern -0.1em V}\xspace}
\newcommand{\ev}{\aunit{e\kern -0.1em V}\xspace}
 
\newcommand{\mevc}{\ensuremath{\aunit{Me\kern -0.1em V\!/}c}\xspace}
\newcommand{\gevc}{\ensuremath{\aunit{Ge\kern -0.1em V\!/}c}\xspace}
\newcommand{\mevcc}{\ensuremath{\aunit{Me\kern -0.1em V\!/}c^2}\xspace}
\newcommand{\gevcc}{\ensuremath{\aunit{Ge\kern -0.1em V\!/}c^2}\xspace}

\def\cm   {\aunit{cm}\xspace}

\newcommand{\chisqndf}{\ensuremath{\chi^2/\mathrm{ndf}}\xspace}

\def\gsim{{~\raise.15em\hbox{$>$}\kern-.85em
          \lower.35em\hbox{$\sim$}~}\xspace}
\def\lsim{{~\raise.15em\hbox{$<$}\kern-.85em
          \lower.35em\hbox{$\sim$}~}\xspace}

\def\tell1  {TELL1\xspace}
\def\ukl1   {UKL1\xspace}

\def\toa{\textrm{ToA}\xspace} 
\def\tot{\textrm{ToT}\xspace} 
\DeclareSIUnit{\clight}{\mathit{c}}
\DeclareSIUnit{\elementarycharge}{\mathit{e}}
\def\timepixfour{\mbox{Timepix4}\xspace}
\def\doca{\mbox{DOCA}\xspace}

\begin{document}
\begin{frontmatter}
\title{First demonstration of \qty{50}{\pico\second} four-dimensional vertex reconstruction with the Timepix4 beam telescope}

\author[a]{D.~Bacher\corref{cor1}} 
\ead{david.bacher@physics.ox.ac.uk}
\author[b]{K.~Akiba} 
\author[b]{M.~van~Beuzekom} 
\author[c]{V.~Coco}  
\author[c]{R.~Dumps} 
\author[b]{T.~Evans} 
\author[b]{K.~Heijhoff} 
\author[a]{M.~John} 
\author[c]{E.~Lemos~Cid} 
\author[c]{L.~Martinazzoli} 
\author[c]{T.~Pajero} 
\author[b]{Y.~Takahashi} 
\author[b]{E.~de~Wit} 

\cortext[cor1]{Corresponding author}

\affiliation[a]{
  organization={Department of Physics, University of Oxford},
  addressline={Denys Wilkinson Bldg., Keble Road},
  city={Oxford},
  postcode={OX1~3RH},
  country={United Kingdom}
}
\affiliation[b]{
  organization={Nikhef},
  addressline={Science Park 105},
  city={Amsterdam},
  postcode={1098 XG},
  country={the Netherlands}
}
\affiliation[c]{
  organization={CERN},
  addressline={Esplanade des Particules 1},
  city={Geneva},
  postcode={1211},
  country={Switzerland}
}

\begin{abstract}
The first demonstration of 4D vertex reconstruction at the \qty{50}{\pico\second} scale using a silicon pixel detector is reported.
Hadronic interactions are produced by directing a \qty[per-mode=symbol]{180}{\giga\electronvolt\per\clight} mixed-hadron beam onto a fixed target upstream of the \timepixfour{} beam telescope at the CERN SPS.
The resulting particles are reconstructed using 4D track and vertex fits that include time directly alongside the spatial parameters.
The vertex time resolution follows the expected statistical scaling with the number of timing measurements, reaching \mbox{\qty{54\pm5}{\pico\second}} for vertices reconstructed from \num{16} timing-plane clusters.
An irreducible component of \mbox{\qty{29.2\pm0.6}{\pico\second}} remains in the fitted resolution model, including sizeable contributions from the clock distribution and the time reference system rather than from the combination of the timing measurements.
The observed scaling down to \qty{54}{\pico\second} without an unforeseen resolution floor above represents a significant experimental step towards precision-timing in future high-pile-up detectors.
\end{abstract}


\end{frontmatter}


\section{Introduction}\label{sec:introduction}
The \lhcb{} \upgradetwo, planned for \lhc{} Run~5, targets a \num{5}-fold increase in instantaneous luminosity to \qty[per-mode=power]{e34}{\per\centi\meter\squared\per\second}~\cite{LHCbcollaboration:2903094}.
At the resulting pile-up of approximately \num{40} interactions per bunch crossing, the spatial overlap of primary collisions makes the correct association of tracks and secondary vertices to their production vertices difficult.
The proposed strategy exploits the temporal spread of approximately \qty{180}{\pico\second} among primary collisions within a bunch crossing~\cite{Albrecht:2653011}, allowing otherwise ambiguous associations to be resolved using both spatial and temporal information.
Simulation studies indicate that a track time resolution of approximately \qty{20}{\pico\second}, corresponding to a hit time resolution of \qty{50}{\pico\second}, is required to obtain the target performance~\cite{LHCbcollaboration:2903094,LHCb-TDR-023}.

Achieving this precision is primarily a detector challenge.
While time resolutions below \qty{50}{\pico\second} have been demonstrated for prototype sensors and readout ASICs developed within the Timespot and IGNITE projects~\cite{Cadeddu:2022kwx,Albicocco:2026xjw,Borgato:2024rft,Lampis:2022lpj}, achieving comparable performance in large-area hybrid assemblies, in which the sensor is bump-bonded to the readout ASIC, has not yet been demonstrated.
The best single-hit time resolutions currently achieved for such assemblies are approximately \qty{130}{\pico\second} with TDCPix~\cite{gigatracker} and \qty{180}{\pico\second} with \timepixfour~\cite{Timepix4Telescope:2025}.
Lower effective time resolutions must therefore be probed by combining multiple timestamps from different detector planes and tracks.
This approach has been demonstrated for beam tracks by the NA62 GigaTracker, which achieved a track time resolution of \qty{65}{\pico\second}~\cite{gigatracker}, and by the \timepixfour{} beam telescope, which reached approximately \qty{90}{\pico\second}~\cite{Timepix4Telescope:2025}.
Vertex reconstruction extends this principle to measurements from multiple tracks, providing a direct way to study the scaling of the precision with the number of measurements and to determine whether correlated contributions limit the achievable resolution.

This paper presents such a study using data obtained with the \timepixfour{} beam telescope~\cite{Timepix4Telescope:2025}.
Hadronic interactions of a mixed-hadron beam in a fixed target are reconstructed, with the resulting particles combined in 4D track and vertex fits, and the spatial and temporal resolution of the reconstructed vertices is studied.
The experimental setup and supporting simulation studies are described in \cref{sec:setup,sec:simulation}, followed by the track and vertex reconstruction procedure in \cref{sec:reconstruction}.
The spatial and temporal vertex resolutions are presented in \cref{sec:spatial_resolution,sec:temporal_resolution}, respectively, before drawing conclusions in \cref{sec:conclusion}.

\section{Setup}\label{sec:setup}
\subsection{Timepix4 beam telescope}\label{sec:setup:telescope}
The Timepix4 beam telescope, described in detail in Ref.~\cite{Timepix4Telescope:2025}, is based on the \timepixfour{} ASIC~\cite{llopart2022timepix4} bump bonded to planar silicon sensors.
The \timepixfour{} ASIC consists of a matrix of \numproduct{512x448} pixels with a pitch of \qtyproduct{55x55}{\micro\meter}.
Each pixel provides a time-of-arrival (\toa) measurement with a nominal binning of \qty{195}{\pico\second} and a time-over-threshold (\tot) measurement with \qty{1.56}{\nano\second} binning.
The timing performance of the bare \timepixfour{} ASIC and bump-bonded assemblies has been extensively characterised in Refs.~\cite{Heijhoff_2022,Bolzonella:2024}.

The telescope consists of eight \timepixfour{} planes arranged in two arms each comprising four planes each.
A right-handed coordinate system is defined with its origin at the centre of the telescope, the $z$-axis along the beam direction and the y-axis pointing upwards.
The four outer planes of the telescope are equipped with \qty{100}{\micro\meter} thick sensors and are positioned perpendicular to the beam to optimise temporal resolution.
The four inner planes are equipped with \qty{300}{\micro\meter} thick sensors and are positioned at an angle of \qty{9}{\degree} around both the $x$- and $y$-axes to optimise spatial resolution at the telescope centre.
The \qty{100}{\micro\meter} thick planes are therefore referred to as \emph{timing planes} and the \qty{300}{\micro\meter} thick planes as \emph{spatial planes}.

For the studies presented in this paper, the telescope was operated at the CERN H8 beamline~\cite{sps-h8} with a \qty{180}{\GeV\per\clight} mixed-hadron beam impinging on a fixed target placed upstream of the telescope.
Data were collected in July and October 2025 and reconstructed using the \textsc{Kepler}\footnote{\url{https://gitlab.cern.ch/TimepixTelescope/Kepler/}} software.

\subsection{Fixed target}\label{sec:setup:target}
A fixed target was installed approximately \qty{12}{\centi\meter} upstream of the front arm to enable vertexing studies.
Secondary particles produced in interactions within the target are reconstructed as tracks and combined into vertices.
Two target configurations are used: a four-layer tungsten target for the longitudinal spatial resolution measurement, and a single-layer brass target with a slit for the transverse spatial resolution measurement.

\begin{figure}
    \centering
    \begin{subfigure}{0.5\textwidth}
        \centering
        \begin{minipage}[c]{0.45\textwidth}
            \centering
            \includegraphics[width=\textwidth]{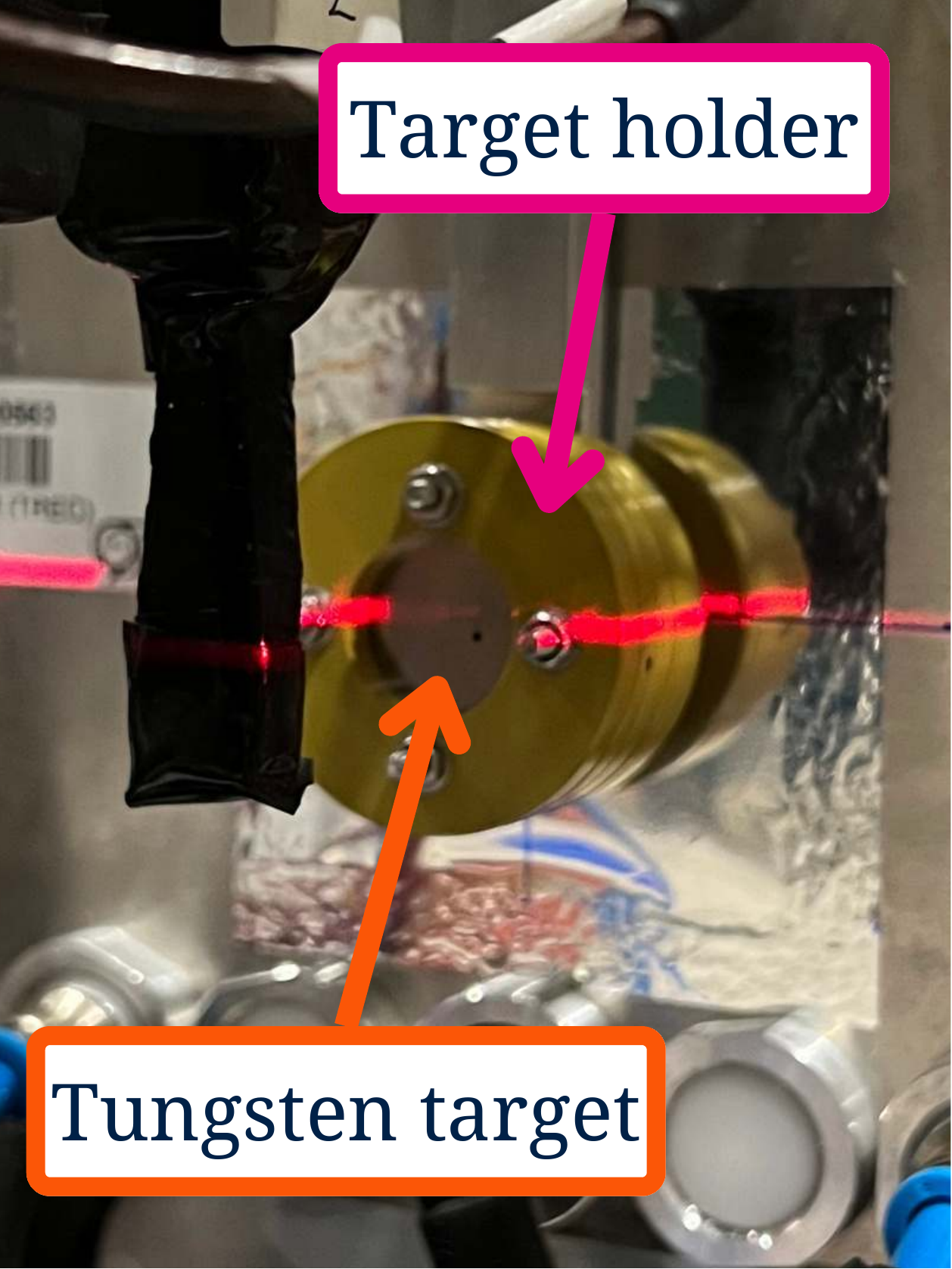}
        \end{minipage}%
        \hspace{0.02\textwidth}%
        \begin{minipage}[c]{0.45\textwidth}
            \centering
            \includegraphics[height=\textwidth,angle=-90]{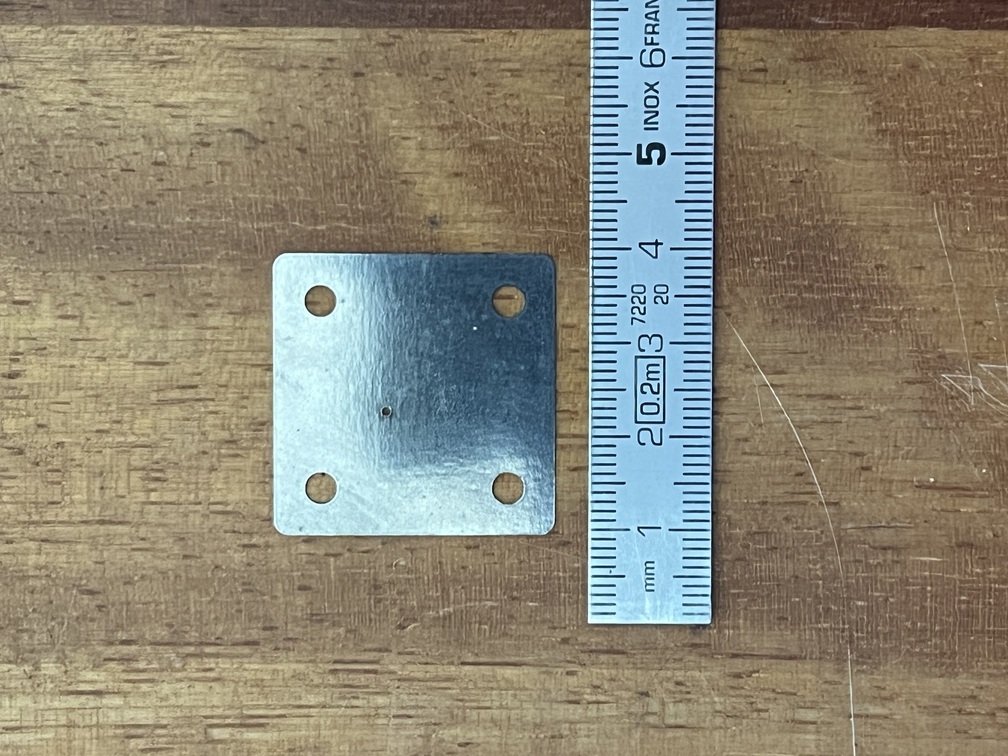}
        \end{minipage}
        \caption{}\label{fig:setup:target:tungsten}
    \end{subfigure}%
    \hfill%
    \begin{subfigure}{0.5\textwidth}
        \centering
        \begin{minipage}[c]{0.45\textwidth}
            \centering
            \includegraphics[width=\textwidth]{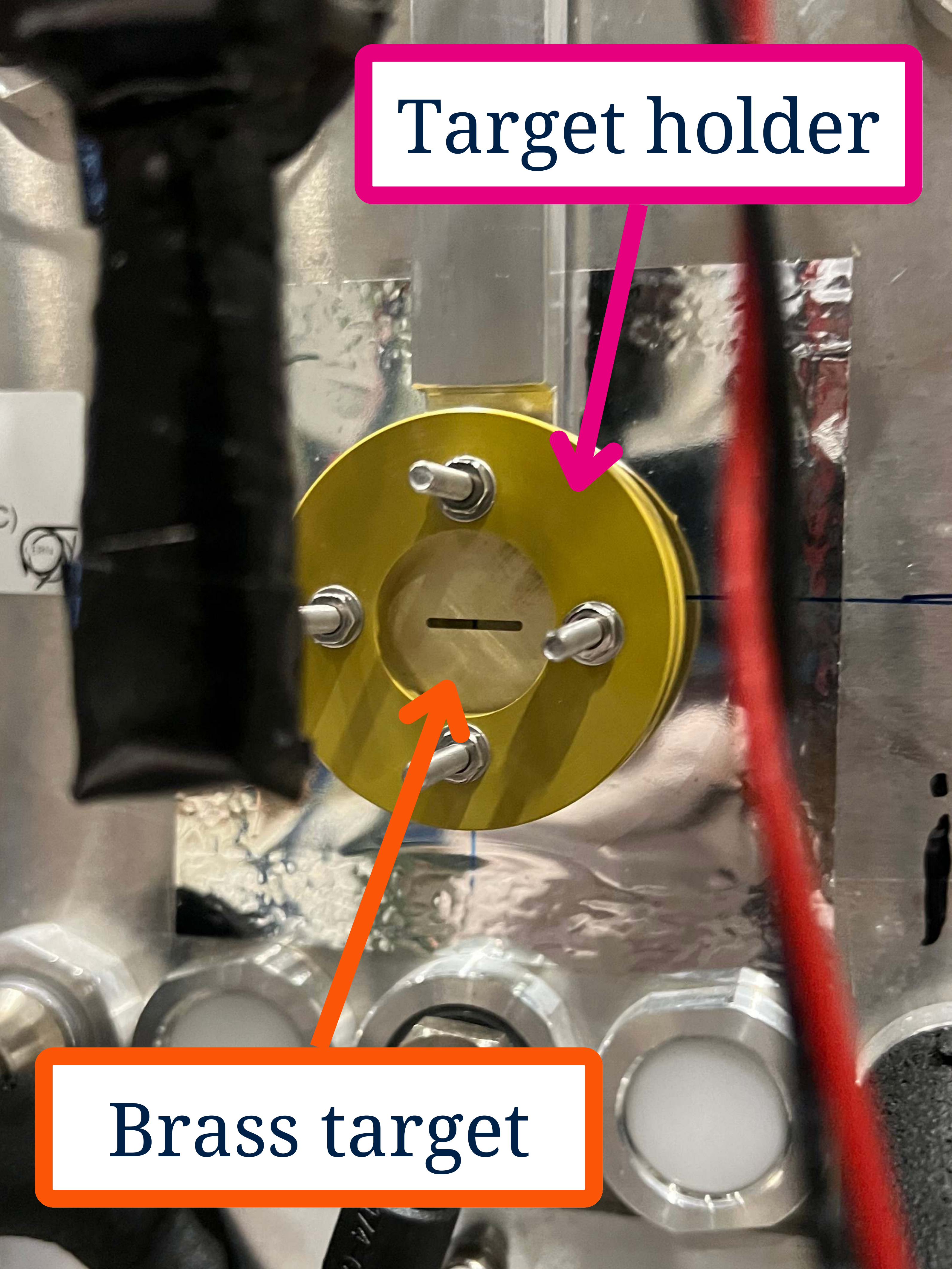}
        \end{minipage}%
        \hspace{0.02\textwidth}%
        \begin{minipage}[c]{0.45\textwidth}
            \centering
            \includegraphics[height=\textwidth,angle=-90]{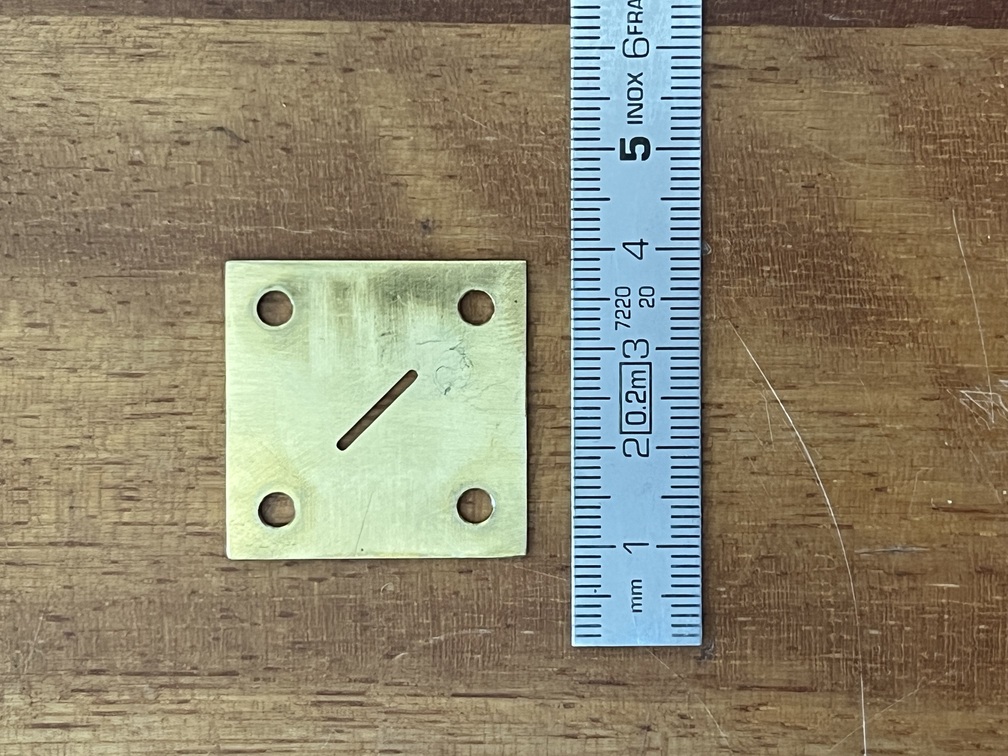}
        \end{minipage}
        \caption{}\label{fig:setup:target:brass}
    \end{subfigure}
    \caption{\subref{fig:setup:target:tungsten} Tungsten target used for the longitudinal spatial resolution measurement, and \subref{fig:setup:target:brass} brass target with the slit used for the transverse measurement, both shown installed in the beamline (left) and in the laboratory (right).
    }\label{fig:setup:targets}
\end{figure}

The tungsten target consists of four layers of \qty{100\pm5}{\micro\meter} thick tungsten separated by \qty{5.04\pm0.03}{\milli\meter}, providing four well-defined interaction planes.
The brass target consists of a single \qty{1000\pm5}{\micro\meter} thick plate containing a slit of width \qty{1130\pm5}{\micro\meter} and length \qty{1}{\cm}.
Data were acquired with the slit orientated along either the $x$- or $y$-axis.
Both targets are shown in \cref{fig:setup:targets}.

For the temporal resolution studies, the tungsten target was installed, but vertices from interactions in any upstream material (the plastic scintillators, the target, and the target holder) were used to maximise the size of the data sample.

\subsection{Time reference system}\label{sec:setup:time_reference}
The telescope is complemented by a time reference system based on two microchannel plate (MCP) detectors, which provide a precise timestamp for each event.
Compared with Ref.~\cite{Timepix4Telescope:2025}, the system was redesigned for use with the fixed-target setup and to reduce the material budget in the beamline.
In the previous downstream configuration, the large angular spread of the secondary particles meant that few traversed the MCPs.
Moreover, the MCPs were placed directly in the beamline, degrading the beam quality and limiting the achievable intensity.

The redesigned system consists of two Photonis PP2365Y\footnote{Photonis Scientific Inc, 660 Main Street, Fiskdale MA, Sturbridge, MA 01566, United States.} MCPs, each coupled to a quartz Cherenkov radiator of dimensions \qtyproduct{16x10x100}{\milli\meter}.
The quartz radiators are positioned upstream of the telescope at an angle of approximately \qty{45}{\degree} to maximise the Cherenkov light yield, while the MCPs are positioned outside the beamline to minimise the material in the beam.
Both the quartz radiators and MCPs are enclosed in a light-tight plastic holder.
The readout and digitisation of the MCP signals are unchanged from Ref.~\cite{Timepix4Telescope:2025}.

\begin{figure}
    \centering
    \includegraphics[width=0.45\textwidth]{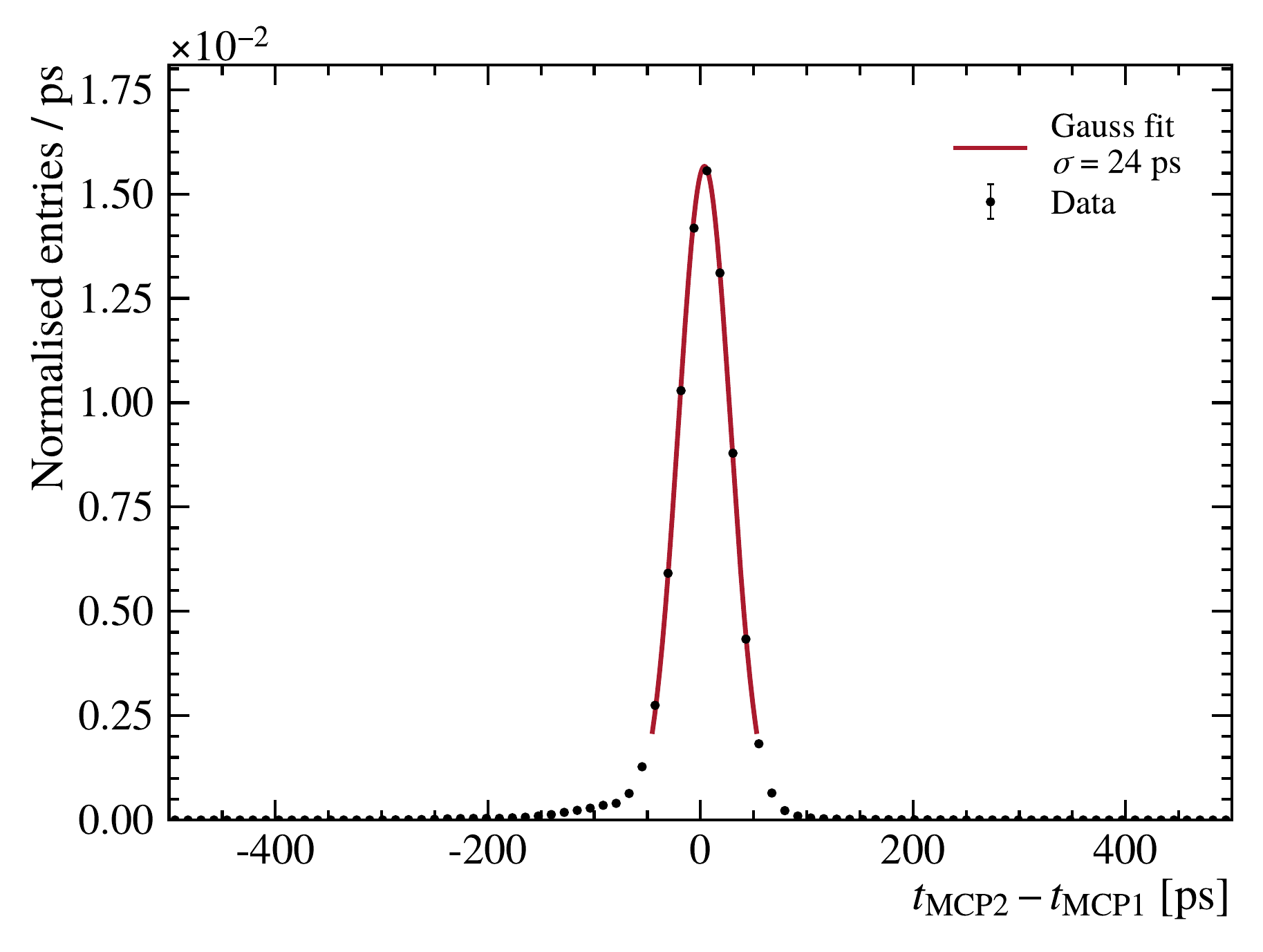}
    \caption{Difference between the timestamps recorded by the two MCPs.
    Only events within $\num{\pm 2}\sigma$ of the main peak are retained for the timing studies.
    }\label{fig:setup:mcp_time_diff}
\end{figure}

The difference between the digitised timestamps of the two MCP signals, as read out from the PicoTDC board, is shown in \cref{fig:setup:mcp_time_diff}.
A Gaussian fit to the distribution yields a standard deviation of \qty{24}{\pico\second}, corresponding to a time resolution of \qty{17}{\pico\second} per MCP and \qty{12}{\pico\second} for the average of the two MCPs (assuming equal performance).
This performance matches that of the central peak in Ref.~\cite{Timepix4Telescope:2025}.
The secondary peak from hadronic interactions observed with the previous system is greatly reduced owing to the smaller material budget of the quartz radiators.
Only events for which the two MCP timestamps agree to within $\qty{\pm48}{\pico\second}$ are retained for the timing studies.

\section{Simulation studies}\label{sec:simulation}
The momentum distribution of secondary particles produced in the fixed-target interactions is characterised using a simulation of the setup implemented in the \textsc{AllPix$^2$}~\cite{allpix2} framework.
The telescope and the fixed target, in either of the configurations described in \cref{sec:setup:target}, are included in the simulation.
Their results are very similar and only those for the tungsten target are presented.
Since the beam at the SPS H8 beamline mostly consists of protons, a proton beam with a kinetic energy of \qty[per-mode=symbol]{180}{\giga\electronvolt} is simulated to impinge on the target, and interactions of the beam particles with the material are modelled using the \textsc{Geant4}~\cite{geant4} software package applying the \texttt{FTFP\_BERT\_LIV} physics model~\cite{Allison:2016lfl}.
Events are retained for analysis if at least two charged particles produced at the vertex pass through the acceptance of the first telescope plane.

\begin{figure}
    \centering
    \begin{subfigure}{0.5\textwidth}
        \centering
        \includegraphics[width=0.9\textwidth]{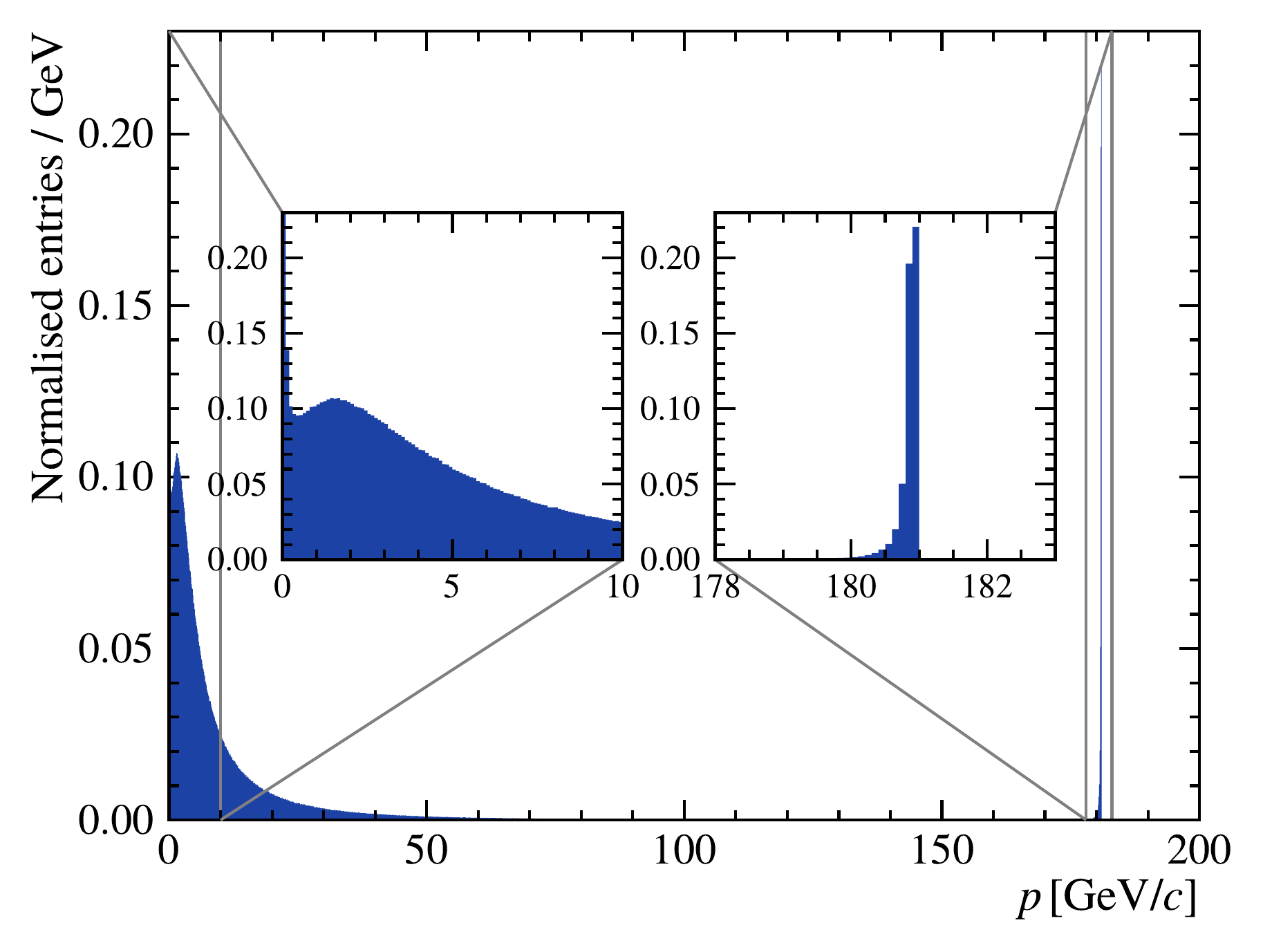}
        \caption{}\label{fig:setup:simulation:momentum:dist}
    \end{subfigure}%
    \hfill%
    \begin{subfigure}{0.5\textwidth}
        \centering
        \includegraphics[width=0.9\textwidth]{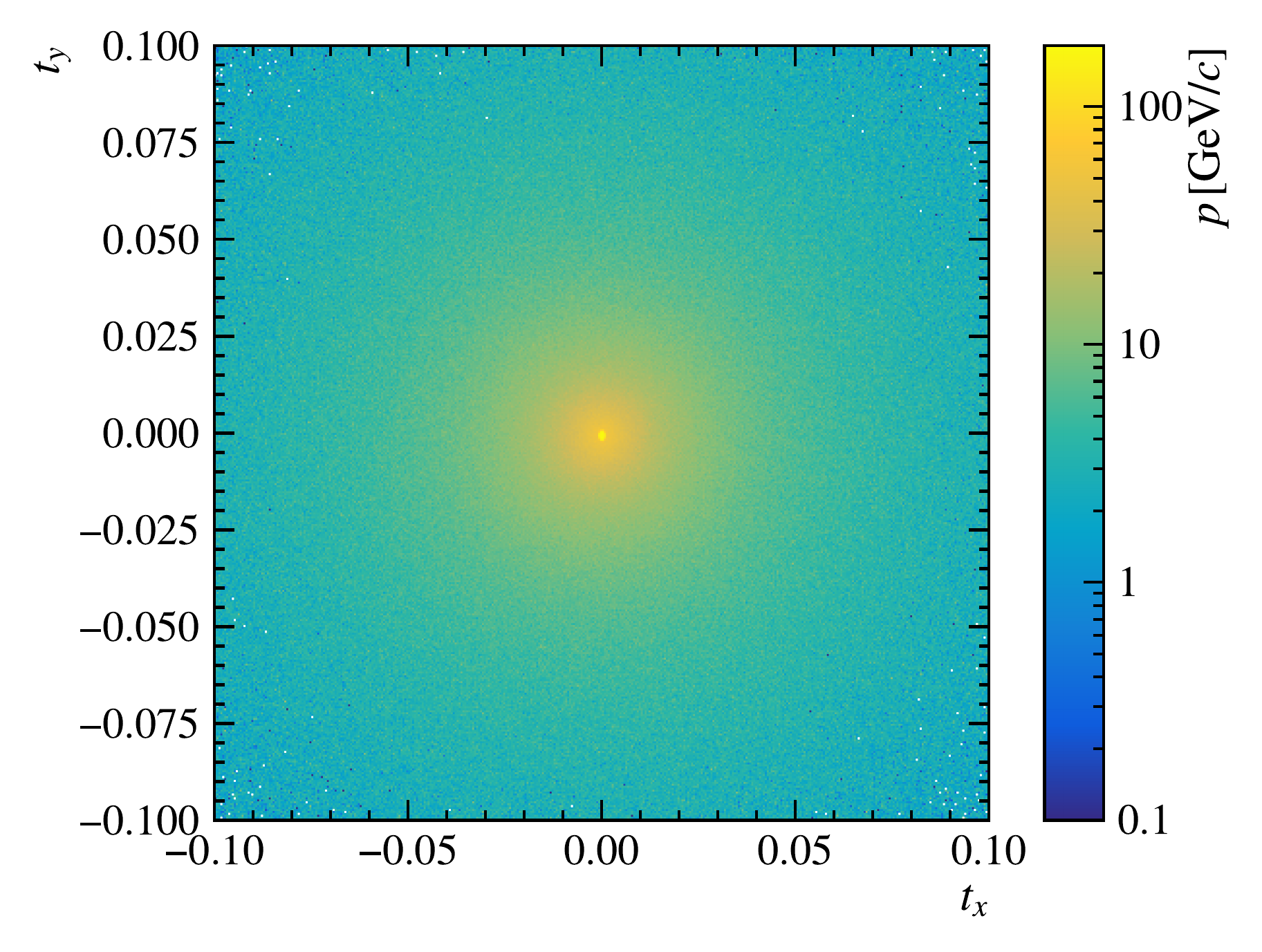}
        \caption{}\label{fig:setup:simulation:momentum:map}
    \end{subfigure}
    \caption{\subref{fig:setup:simulation:momentum:dist} Momentum distribution of simulated particles produced at the vertex and within the acceptance of the first telescope plane.
    \subref{fig:setup:simulation:momentum:map} Mean simulated momentum as a function of the track slopes $t_x$ and $t_y$, used to provide momentum estimates for the Kalman filter--based track fit.
    }\label{fig:setup:simulation:momentum}
\end{figure}

The momentum distribution of particles within the acceptance of the first telescope plane is shown in \cref{fig:setup:simulation:momentum:dist}.
It is bimodal, with a narrow peak at the beam momentum of \qty{181}{\giga\electronvolt\per\clight} from primary beam particles undergoing soft interactions, and a broader distribution peaking near \qty{1.5}{\giga\electronvolt\per\clight} from hard interactions.
In practice, only particles from hard interactions are reconstructed, as the low-momentum products of soft interactions are emitted at large angles and fail to traverse the minimum of four planes required for track reconstruction.
The resulting track momenta, typically below \qty{50}{\giga\electronvolt\per\clight}, lead to significant multiple scattering.
These multiple-scattering effects are incorporated using a Kalman filter--based track fit~\cite{Kalman:1960mft}.
The Kalman filter requires an estimate of each track's momentum to determine the corresponding scattering uncertainty.
However, a direct measurement of the momentum is not available, as the telescope does not include a magnetic field.
The momentum is therefore estimated from a finely grained 2D histogram of the mean simulated momentum as a function of the track slopes $t_x$ and $t_y$, shown in \cref{fig:setup:simulation:momentum:map}, evaluated using the slopes from an initial straight-line fit.
All particles are treated as pions, which are found to be the most abundant species in the simulation.

\begin{figure}
    \centering
    \includegraphics[width=0.9\textwidth]{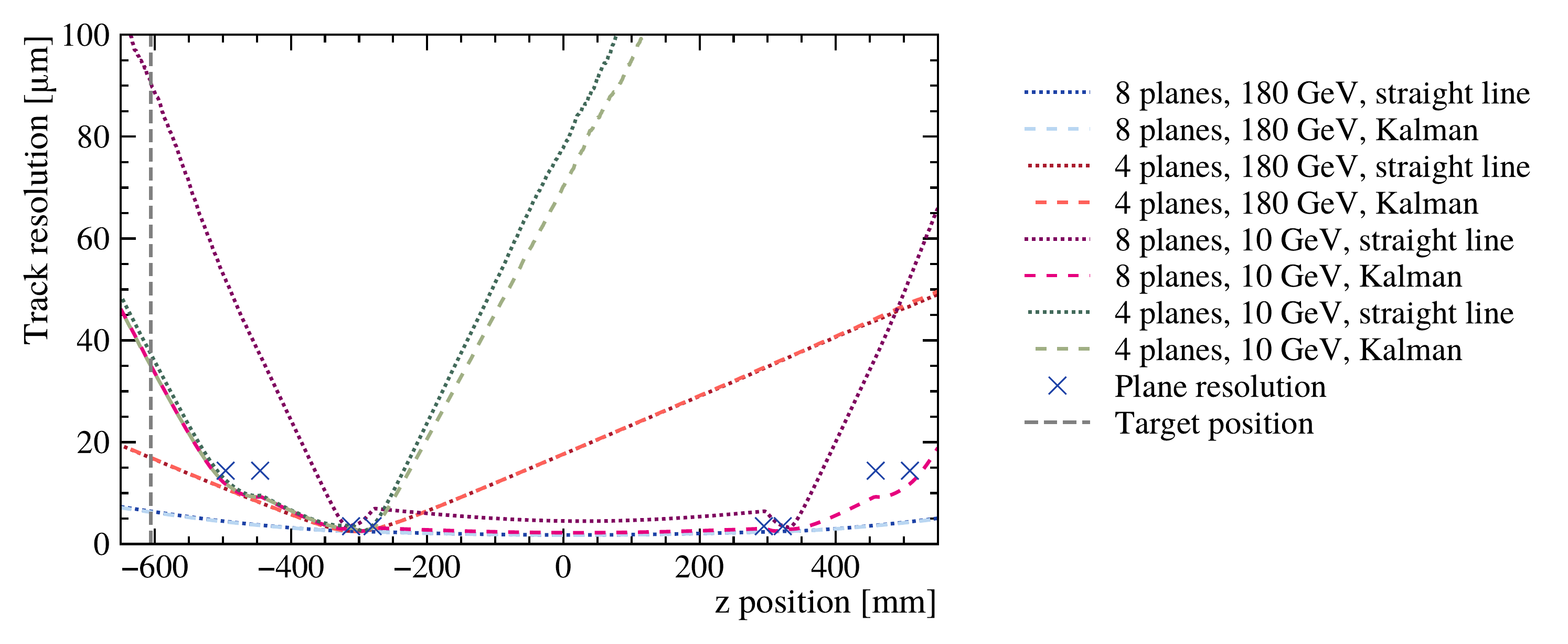}
    \caption{Pointing resolution as a function of $z$ for tracks of \qty{180}{\giga\electronvolt\per\clight} and \qty{10}{\giga\electronvolt\per\clight}, reconstructed with a straight-line or Kalman fit using either all eight planes or only the four front planes.
    The dashed grey line indicates the target position.
    }\label{fig:setup:simulation:track_resolution}
\end{figure}

A fast simulation of the telescope, described in Ref.~\cite{Timepix4Telescope:2025}, is used to study the expected track pointing resolution at the target position.
Straight-line and Kalman filter fits using either four or eight planes are compared for track momenta of \qty{10}{\giga\electronvolt\per\clight}, typical of secondary particles, and \qty{180}{\giga\electronvolt\per\clight}, corresponding to the beam momentum.
The results, shown in \cref{fig:setup:simulation:track_resolution}, confirm that the Kalman filter significantly improves the resolution at low momentum, particularly when all eight planes are used, reaching approximately \qty{35}{\micro\meter} at the target for \qty{10}{\giga\electronvolt\per\clight} tracks.

\section{Reconstruction}\label{sec:reconstruction}
\subsection{Calibrations}
Several calibrations are applied to the raw data from the \timepixfour{} telescope before reconstruction, including charge calibration, $\eta$-corrections, and timing corrections.
The charge calibration, $\eta$-corrections, and frequency corrections of the voltage controlled oscillators (VCOs) of the Timepix4 ASIC follow the procedure described in Ref.~\cite{Timepix4Telescope:2025} and are not discussed further.

The timing calibration procedure has been updated for this study.
The timewalk correction employed in Ref.~\cite{Timepix4Telescope:2025} used pixel-by-pixel timewalk fits.
This method is impractical for this study because the angular spread of the secondary particles illuminates the entire pixel matrix on every plane, requiring a separate calibration for each of the \num{229376} pixels on each plane.
A neural network--based method, described in detail in Ref.~\cite{nn_timewalk}, is therefore used and briefly summarised here.
Instead of fitting a timewalk function to the measured timewalk for each pixel, a feedforward multilayer perceptron (MLP) is trained to predict the timewalk correction from the charge of the earliest hit in the cluster, the track intercept in the local $x$- and $y$-coordinates of the plane (along columns and rows) relative to the centre of that hit.
Compared to Ref~\cite{nn_timewalk} the track slopes $t_x$ and $t_y$ have been added as additional input features.
With this method, a single MLP is trained for each plane, while pixel-to-pixel variations are absorbed into the per-pixel time offsets.
As shown in Ref.~\cite{nn_timewalk}, its performance matches that of pixel-by-pixel timewalk fits for large datasets and exceeds it when the available data are limited.

\subsection{Track and vertex reconstruction}
The track reconstruction is adapted from the standard procedure described in Ref.~\cite{Timepix4Telescope:2025} to accommodate the large angular spread of secondary particles produced at the vertex.
The requirement on the maximum allowed opening angle between the track and the beam axis is removed.
The minimum number of clusters required per track is reduced from eight to four, allowing tracks that traverse only the front arm of the telescope to be retained.
The limit on the number of simultaneous clusters per plane is also removed because multiple coincident tracks are expected from fixed-target interactions.
Tracks are fitted with the Kalman filter implemented in the \textsc{Kepler} reconstruction framework following Ref.~\cite{FRUHWIRTH1987444}, which accounts for multiple scattering in the telescope planes.
The track fit is seeded using a straight-line fit to the clusters, and the track momentum is estimated from a parametrisation of the mean simulated momentum as a function of the track slopes $t_x$ and $t_y$, as described in \cref{sec:simulation}.
The track time is included directly in the track state and fitted simultaneously with the spatial position and slopes, using the VCO- and timewalk-corrected cluster times as measurements.

Tracks within \qty{1}{\nano\second} of one another are combined into vertex candidates by calculating the distance of closest approach (\doca) for every pair and grouping those with a \doca below \qty{200}{\micro\meter}.
In this way, vertices containing up to eight tracks are reconstructed from the data sample.
The vertex position and time, together with the track slopes at the vertex, are then determined using the iterative least-squares fitting procedure of Ref.~\cite{Strandlie:2010zz}.
As in the track fit, time is included directly in the vertex state and fitted simultaneously with the spatial position and track slopes, using the fitted track states as measurements.
The complete procedure therefore constitutes 4D track and vertex fits, rather than a spatial fit to which timing information is added afterwards.

An overview of the standard selection criteria applied to tracks and vertices is given in \cref{tab:reconstruction:selection}.
\begin{table}
    \centering
    \caption{Standard selection requirements for vertex reconstruction.}\label{tab:reconstruction:selection}
    \small
    \begin{tabularx}{\textwidth}{p{4cm}lX}
        \toprule
        Requirement & Default value & Description \\
        \midrule
        Clustering \\
        \midrule
        Time window & \qty{100}{\nano\second} & Maximum time difference between hits in the same cluster \\
        Cluster width in $x$ and $y$ & $\leq \num{2}$ & Rejects large clusters from $\delta$-rays, nuclear interactions, etc. \\
        \multirow{2}{=}{Charge threshold} & $\sim \qty{4}{\kilo\elementarycharge^{-}}$ (timing planes) & \multirow{2}{=}{Suppresses noise, optimised per plane} \\
        & $\sim \qty{12}{\kilo\elementarycharge^{-}}$ (spatial planes) & \\
        \midrule
        Tracking \\
        \midrule
        Time window & \qty{5}{\nano\second} & Maximum time difference between clusters to form a track seed (corrected for time of flight) \\
        Number of clusters per track & $\geq \num{4}$ & Requires clusters on at least four telescope planes \\
        Fit \chisqndf & $\leq \num{30}$ & Rejects poorly reconstructed tracks \\
        \midrule
        Vertexing \\
        \midrule
        Time window & \qty{1}{\nano\second} & Maximum time difference between tracks to form a vertex seed (corrected for time of flight) \\
        \doca & $\leq \qty{200}{\micro\meter}$ & Maximum distance between track pairs to form a vertex seed \\
        Fit \chisqndf & $\leq \num{10}$ & Rejects poorly reconstructed vertices \\
        \bottomrule
    \end{tabularx}
\end{table}

\begin{figure}
    \centering
    \includegraphics[width=0.9\textwidth]{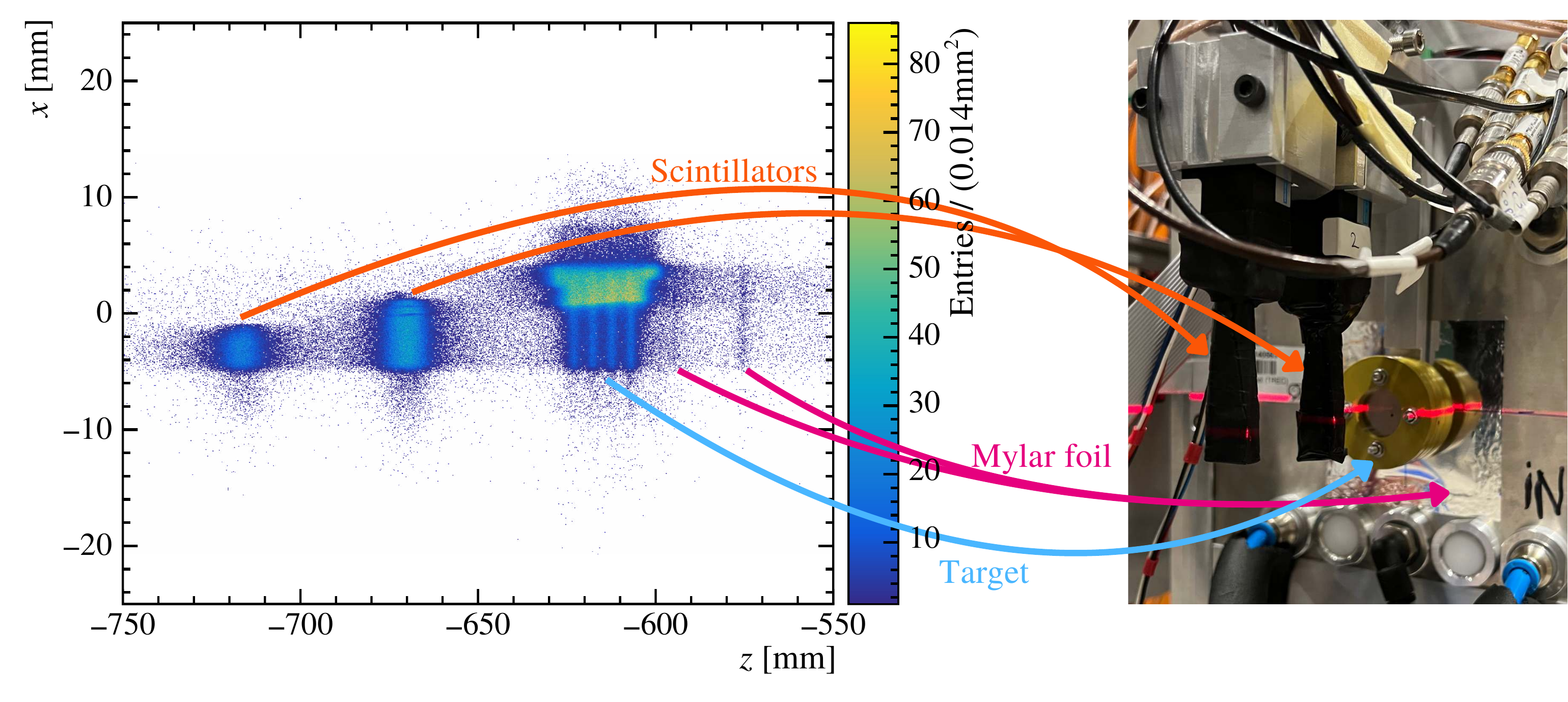}
    \caption{Reconstructed vertex distribution in the $x$-$z$ plane for the tungsten target, with the resolved structural features annotated on the right.
    }\label{fig:reconstruction:vertex_distribution}
\end{figure}

The distribution of reconstructed vertices in the $x$-$z$ plane for the tungsten target configuration is shown in \cref{fig:reconstruction:vertex_distribution}.
The four tungsten layers, upstream plastic scintillators, target holder, and downstream Mylar entrance-window foils of the front-arm telescope enclosure are all clearly resolved.
The target holder is visible because the beam is not perfectly centred on the target.

\begin{figure}
    \centering
    \includegraphics[width=0.45\textwidth]{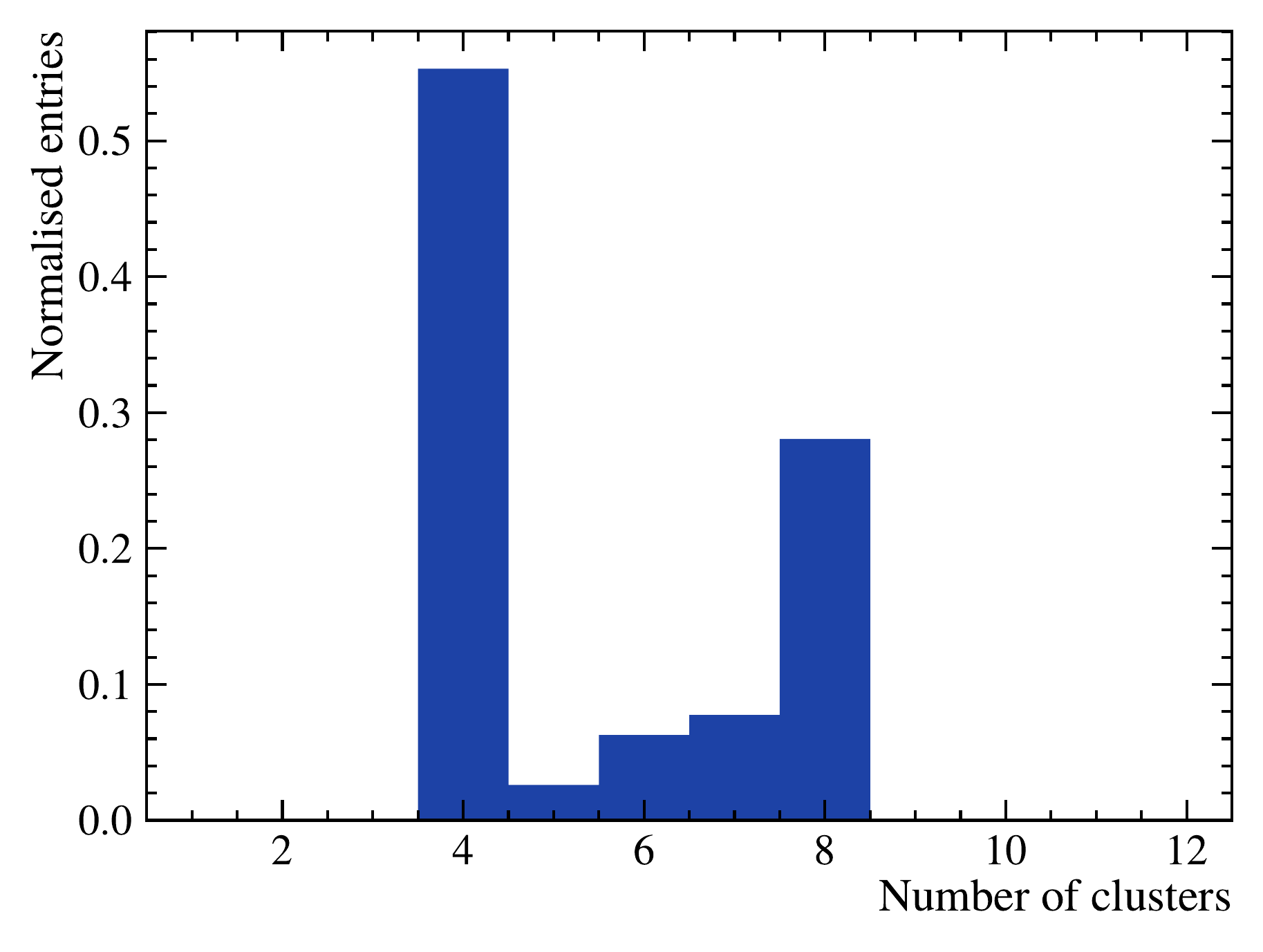}
    \caption{Distribution of the number of clusters per track for tracks used to reconstruct vertices.
    }\label{fig:reconstruction:track_size}
\end{figure}

The distribution of the number of clusters comprising the reconstructed vertex tracks, that is the set of tracks fitted to a single vertex, is shown in \cref{fig:reconstruction:track_size}.
Most tracks produced in fixed-target interactions traverse only the front arm of the telescope and therefore contain four clusters, while a smaller fraction traverse the entire telescope and contain eight.
Tracks with five to seven clusters traverse the front arm and at least one plane of the back arm, but either leave the telescope acceptance or miss one or more planes because of detector inefficiencies.

\section{Spatial resolution}\label{sec:spatial_resolution}
The longitudinal and transverse spatial resolutions are measured using the tungsten and brass targets, respectively.
Although tracks within a vertex may contain different numbers of clusters, multiple scattering limits the dependence of the pointing resolution at the target on the number of clusters in the back arm, as shown in \cref{fig:setup:simulation:track_resolution}.
The vertex resolution is therefore expected to depend primarily on the number of tracks rather than the number of clusters per track, and is studied as a function of the vertex track multiplicity $N_{\text{tracks}}$.

\subsection{Longitudinal resolution}\label{sec:spatial_resolution:longitudinal}
\begin{figure}
    \centering
    \includegraphics[width=0.45\textwidth]{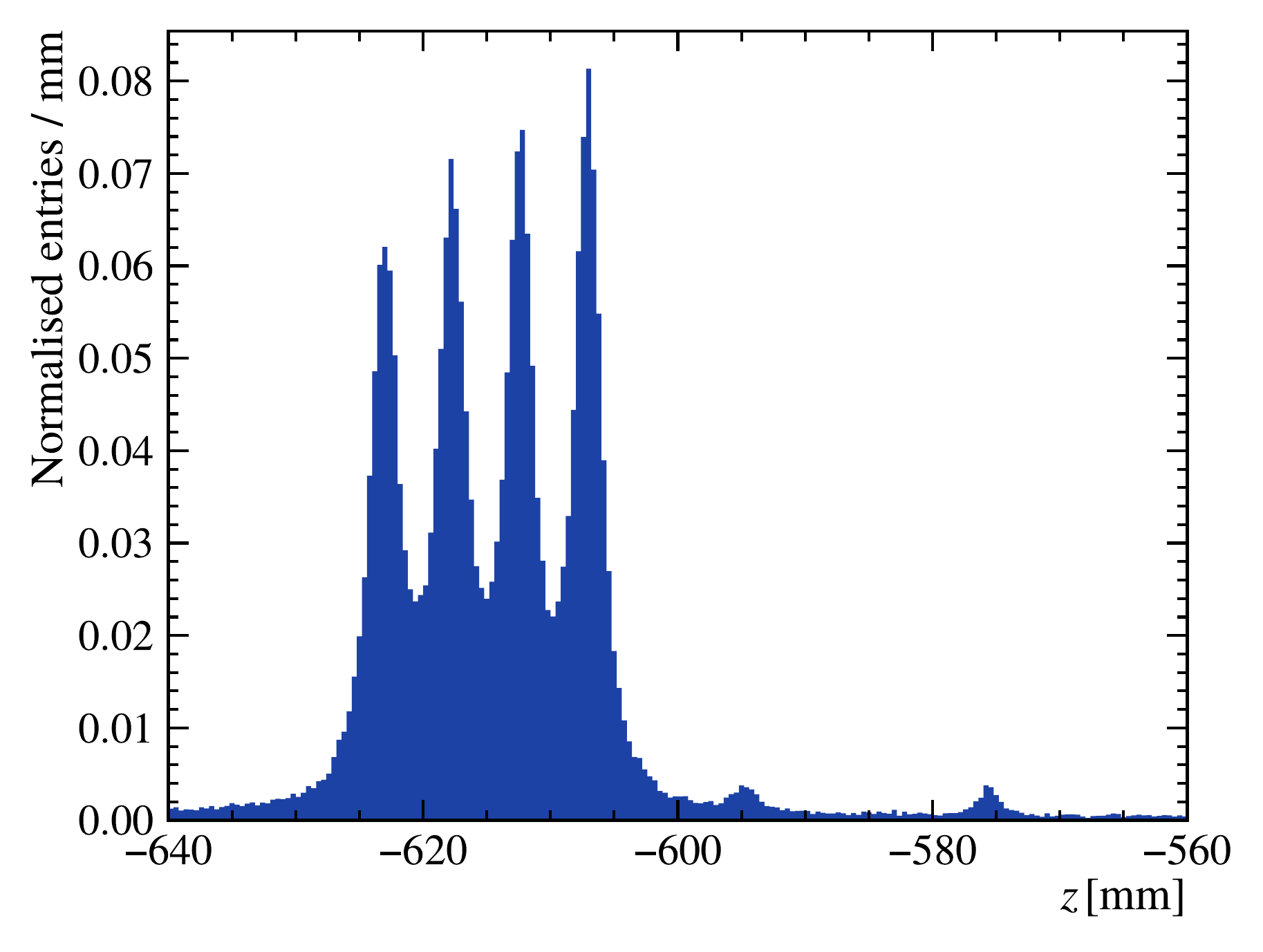}
    \caption{Distribution of reconstructed vertex positions along $z$ for the tungsten target, showing four peaks from the tungsten layers and two smaller peaks from the Mylar entrance-window foils.
    }\label{fig:spatial_resolution:longitudinal:vertex_z}
\end{figure}

The longitudinal spatial resolution is determined using the tungsten target, which provides four clearly defined interaction planes.
The distribution of reconstructed vertex positions along the beam axis, $z$, is shown in \cref{fig:spatial_resolution:longitudinal:vertex_z}.
The four prominent peaks between $z = \qty{-630}{\milli\meter}$ and $z = \qty{-600}{\milli\meter}$ correspond to the four tungsten layers.
Two smaller peaks at $z = \qty{-595}{\milli\meter}$ and $z = \qty{-575}{\milli\meter}$ are also visible, corresponding to the Mylar foils covering the entrance window of the front arm enclosure.
Only vertices reconstructed within $\qty{-630}{\milli\meter} < z < \qty{-600}{\milli\meter}$ are considered for the longitudinal resolution measurement.
Because the target layers may not be perfectly perpendicular to the $z$-axis, an alignment correction is applied by minimising the measured longitudinal resolution with respect to small rotations of the coordinate frame about all three axes.
The rotations are centred on the first target layer.
The resulting coordinate is denoted as $z_{\text{cal}}$.

\begin{figure}
    \centering
    \begin{subfigure}{0.5\textwidth}
        \centering
        \includegraphics[width=0.9\textwidth]{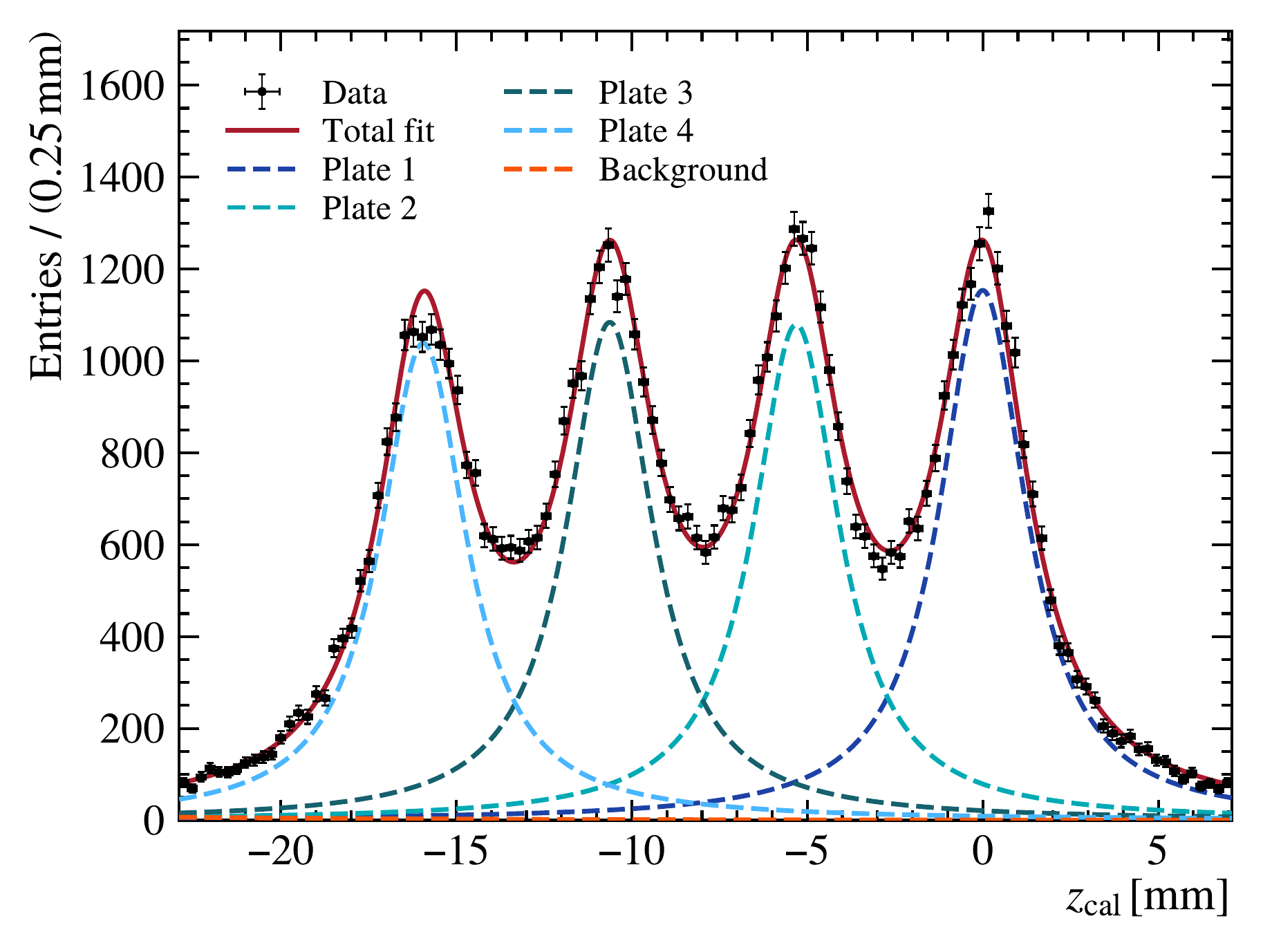}
        \caption{}\label{fig:spatial_resolution:longitudinal:fit:2tracks}
    \end{subfigure}%
    \hfill%
    \begin{subfigure}{0.5\textwidth}
        \centering
        \includegraphics[width=0.9\textwidth]{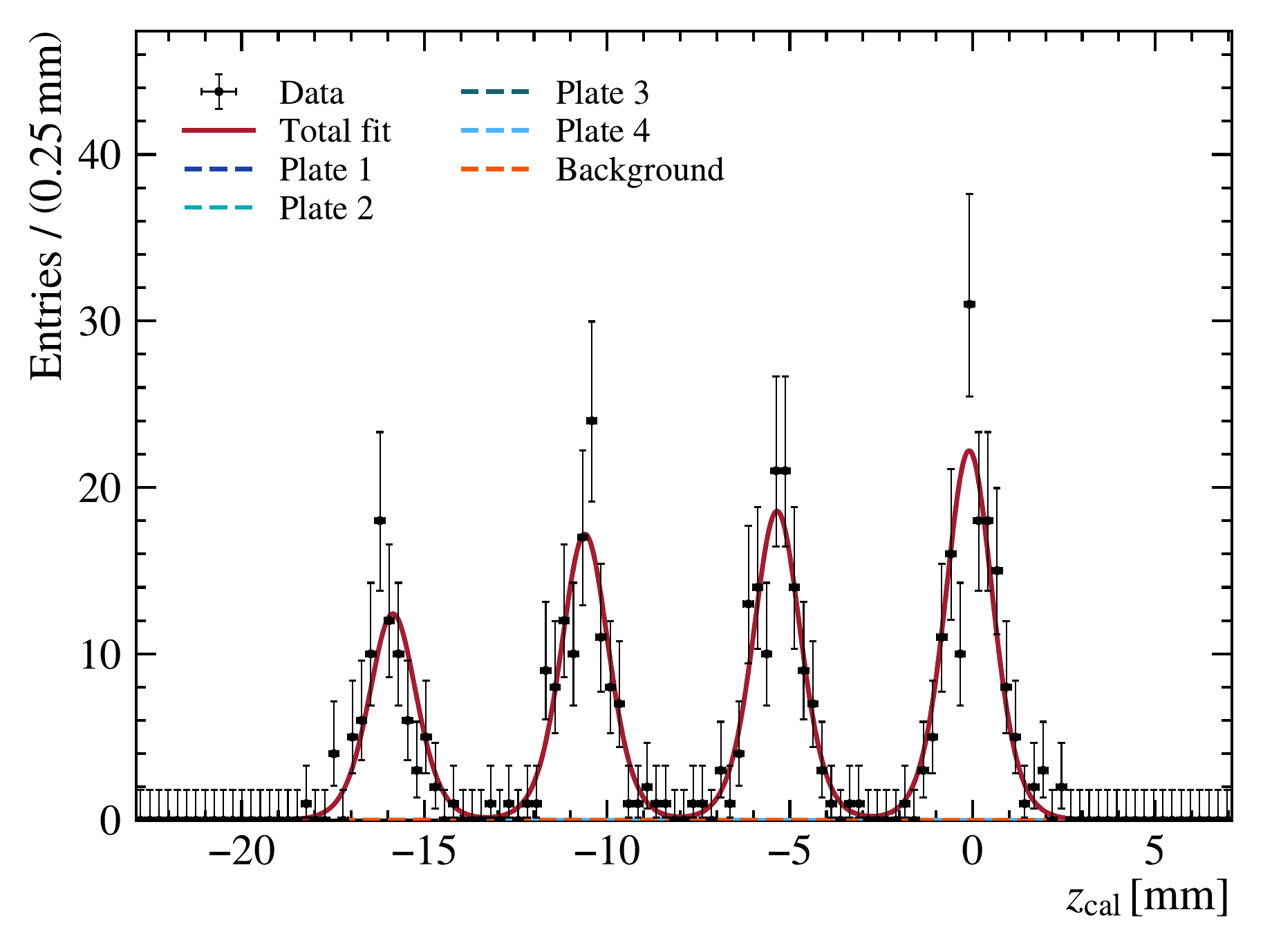}
        \caption{}\label{fig:spatial_resolution:longitudinal:fit:6tracks}
    \end{subfigure}
    \caption{Fits to the $z_{\text{cal}}$ distribution of vertices with \subref{fig:spatial_resolution:longitudinal:fit:2tracks} two and \subref{fig:spatial_resolution:longitudinal:fit:6tracks} six tracks.
    }\label{fig:spatial_resolution:longitudinal:fit}
\end{figure}

The longitudinal resolution is extracted for vertex track multiplicities from two to six using unbinned maximum-likelihood fits to the $z_{\text{cal}}$ distributions.
Each tungsten layer is modelled by a uniform distribution $\mathcal{U}(c_i, d)$ of width $d$ centred at $c_i$, convolved with a Student's t-distribution $t_\nu(\sigma_z)$ with mean zero and scale $\sigma_z$.
The number of degrees of freedom, $\nu = 2N_{\text{tracks}} - 3$, is fixed to that of the vertex fit.
The full signal model is the sum of four such convolutions with a common $\sigma_z$.
The layer thickness $d$ is Gaussian-constrained to $\qty{100\pm5}{\micro\meter}$, and the layer spacings to $\qty{5.04\pm0.03}{\milli\meter}$.
The normalisation of each layer and an exponential background, which describes the small contribution from non-tungsten interactions and fake vertices, are free parameters.
Example fit results for vertices with two and six tracks are shown in \cref{fig:spatial_resolution:longitudinal:fit}.

\begin{figure}
    \centering
    \includegraphics[width=0.45\textwidth]{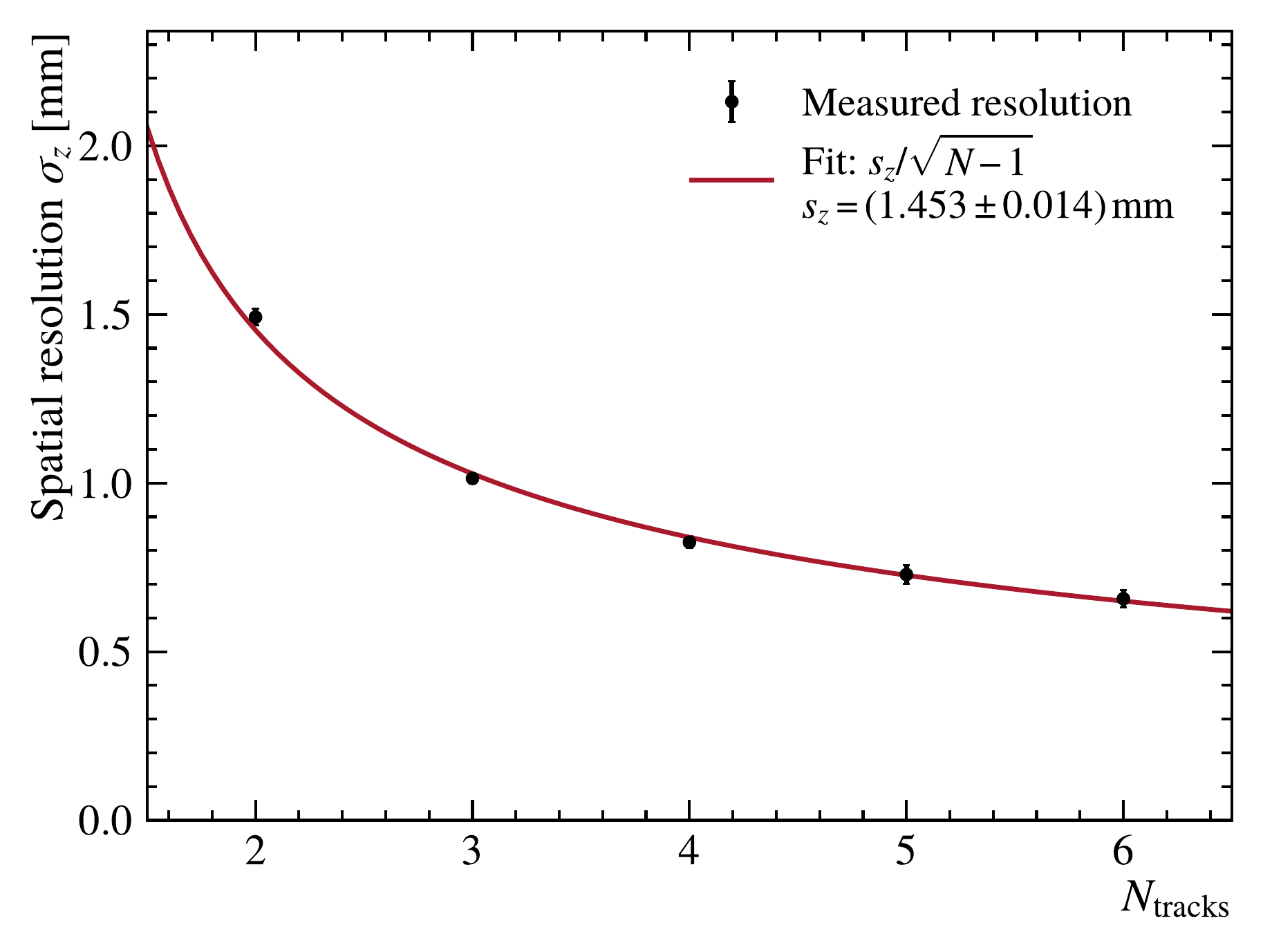}
    \caption{Longitudinal resolution $\sigma_z$ as a function of $N_{\text{tracks}}$, with the best fit to $\sigma_z = s_z / \sqrt{N_{\text{tracks}} - 1}$ overlaid.
    The uncertainties shown have been scaled by the Birge ratio $\sqrt{\chisqndf} \approx \num{2.16}$.
    }\label{fig:spatial_resolution:longitudinal:resolution}
\end{figure}

The extracted longitudinal resolution $\sigma_z$ is shown as a function of the track multiplicity, $N_{\text{tracks}}$, in \cref{fig:spatial_resolution:longitudinal:resolution}.
The resolution improves with track multiplicity, reaching $\sigma_z = \qty{0.657\pm0.025}{\milli\meter}$ for vertices with six tracks.
The dependence on track multiplicity is well described by
\begin{equation}
    \sigma_z(N_{\text{tracks}}) = \frac{s_z}{\sqrt{N_{\text{tracks}} - 1}},
\end{equation}
where the denominator reflects the fact that a single track does not constrain the $z$-position of the vertex.
The fit yields $s_z = \qty{1.453\pm0.014}{\milli\meter}$ with $\chisqndf = \num{4.6}$, suggesting that the uncertainties do not fully account for systematic effects.
The uncertainties on the measurements shown in \cref{fig:spatial_resolution:longitudinal:resolution}, and on $s_z$ are therefore scaled by the  Birge ratio $\sqrt{\chisqndf} \approx \num{2.16}$~\cite{Birge:1932hlj}.

The use of a Student's t-distribution rather than a Gaussian is motivated by the imperfect calibration of the track covariances, as such a distribution arises naturally when the variance of an otherwise Gaussian distributed variable is uncertain.
Two sources of covariance miscalibration are expected.
First, the per-track momentum estimates are derived from a slope parametrisation of the mean momentum rather than direct measurements, introducing uncertainty into the estimated multiple-scattering contributions.
Second, each telescope plane has a titanium cooling block on its rear surface with a circular central cutout for the beam.
Tracks passing through the corners of a plane therefore traverse more material than those near its centre, but this variation is not included in the track fit.
The effect is amplified in $z$, where the small angles between tracks cause modest angular errors to produce large uncertainties in their intersection point.

\subsection{Transverse resolution}
\begin{figure}
    \centering
    \includegraphics[width=0.45\textwidth]{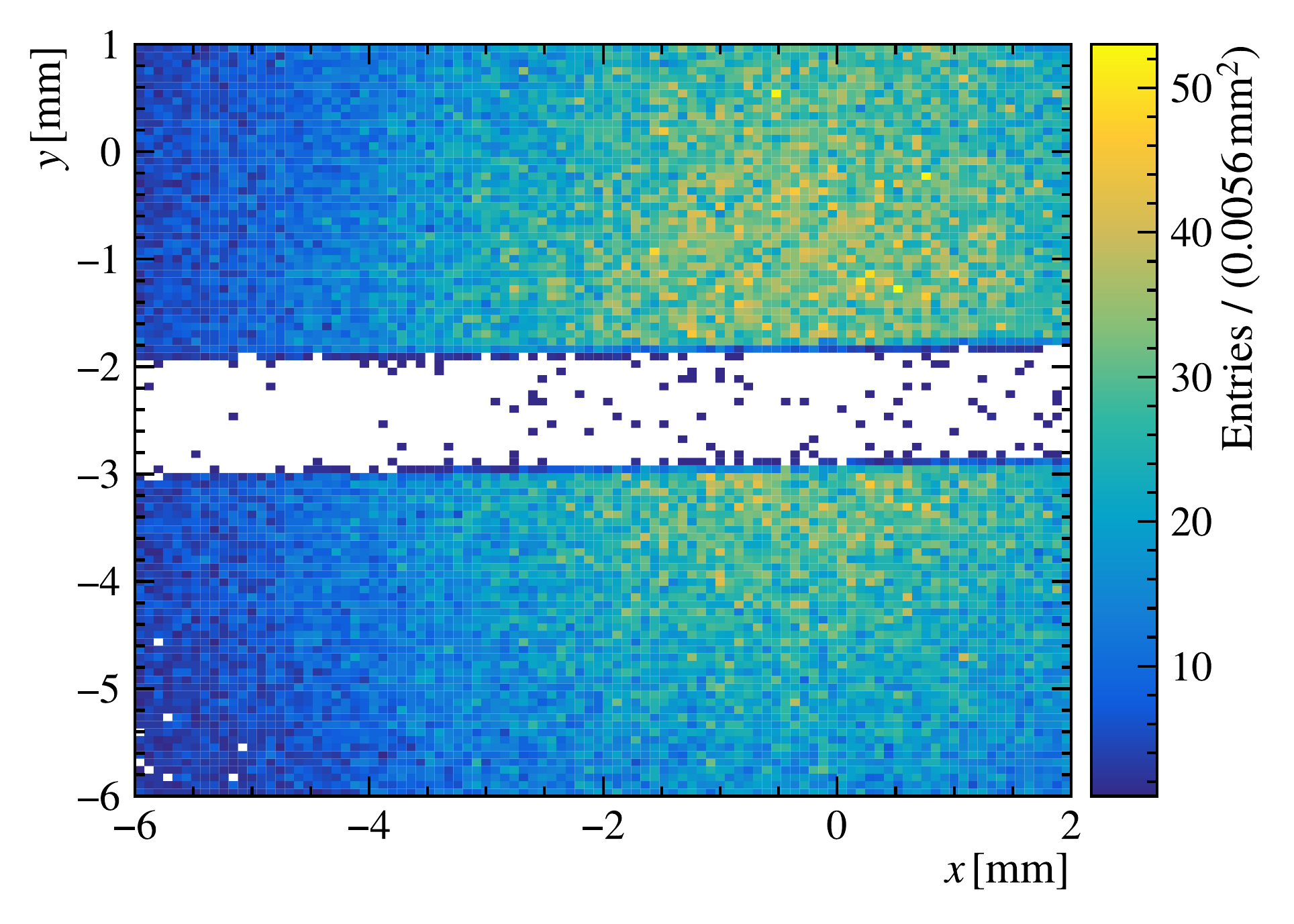}
    \caption{Vertex distribution in the transverse plane for the brass target with the slit oriented along $x$.
    }\label{fig:spatial_resolution:transverse:vertex_xy}
\end{figure}

The transverse spatial resolution is studied using the brass target, whose slit provides a sharp edge in the transverse plane.
Data were taken with the slit oriented along both the $x$ and $y$ directions, enabling the resolution to be measured independently in each transverse coordinate.
The distribution of reconstructed vertex positions for the $x$-oriented slit is shown in \cref{fig:spatial_resolution:transverse:vertex_xy}, where the slit appears as a band of low vertex density between $y = \qty{-3}{\milli\meter}$ and $y = \qty{-2}{\milli\meter}$.
An alignment procedure analogous to that described in \cref{sec:spatial_resolution:longitudinal} yields calibrated coordinates $x_{\text{cal}}$ and $y_{\text{cal}}$.
The transverse resolution is then extracted for $N_{\text{tracks}}$ between two and five using unbinned maximum-likelihood fits.

Since the vertex distribution is modulated by the primary beam profile, a kernel density estimate $\mathcal{B}(x_{\text{cal}})$ obtained from non-interacting beam tracks is included as a multiplicative term in the fit model.
The target slit is modelled as a step function of width $w$ centred at $c$, convolved with a Gaussian with standard deviation $\sigma_x$.
The slit width $w$ is Gaussian-constrained to \qty{1.130\pm0.005}{\milli\meter}.
To account for a non-uniform geometrical acceptance of the vertices, a linear acceptance term $1 + \beta x_{\text{cal}}$ is included, with the slope $\beta$ free in the fit.
The full model is thus given by
\begin{align}
    \mathcal{P}_s(x_{\text{cal}}; c, w, \sigma_x, \beta) &= \mathcal{B}(x_{\text{cal}}) \cdot \left(1 + \beta x_{\text{cal}}\right)\nonumber\\
    &\quad \cdot \left[ \left(1 - w\,\mathcal{U}(c, w) \right) * \mathcal{N}(0, \sigma_x)\right](x_{\text{cal}}).
\end{align}
A uniform background accounts for the small number of vertices reconstructed outside the target and for fake vertices.

\begin{figure}
    \centering
    \begin{subfigure}{0.5\textwidth}
        \centering
        \includegraphics[width=0.9\textwidth]{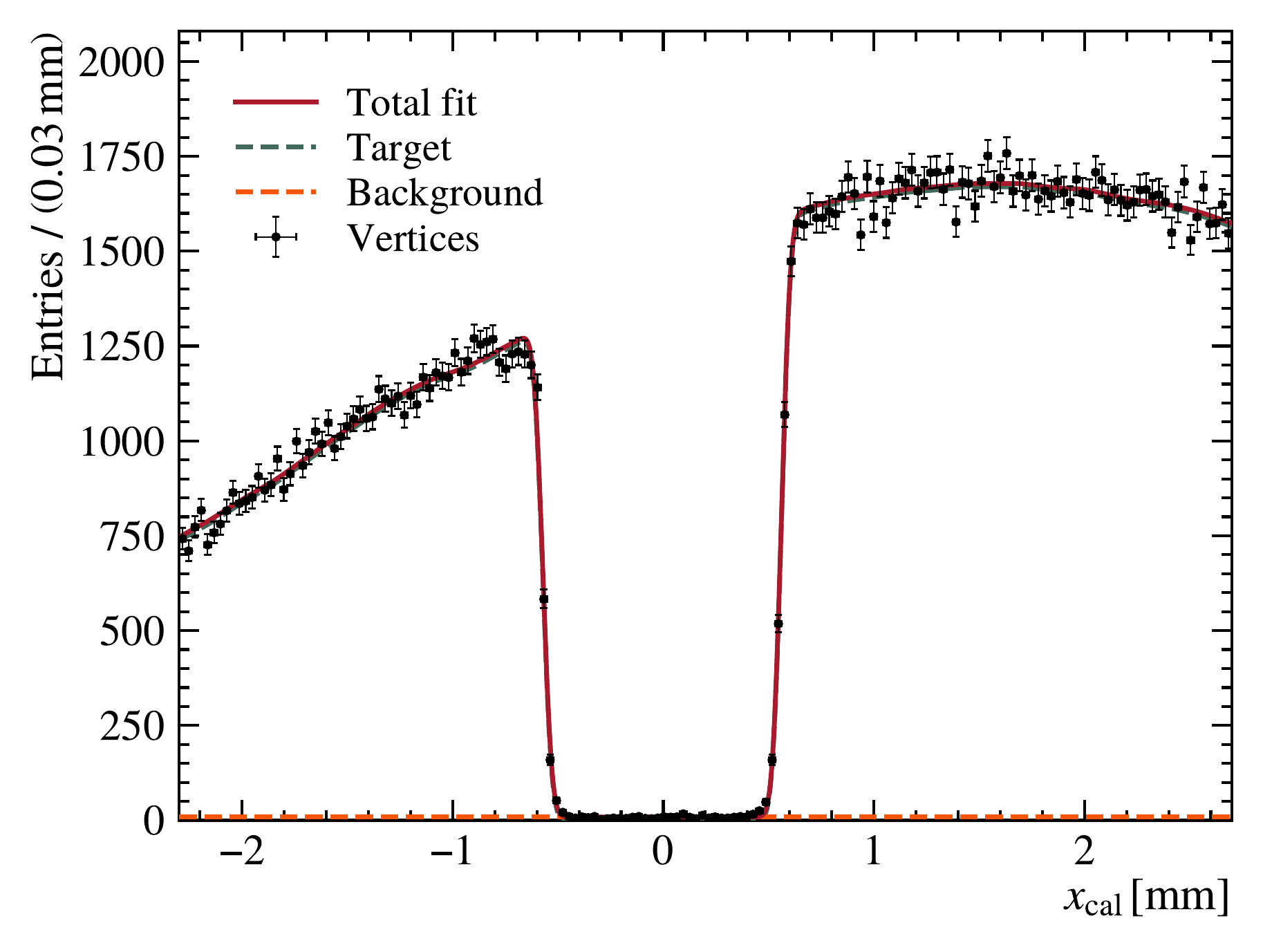}
        \caption{}\label{fig:spatial_resolution:transverse:fit:x2tracks}
    \end{subfigure}%
    \hfill%
    \begin{subfigure}{0.5\textwidth}
        \centering
        \includegraphics[width=0.9\textwidth]{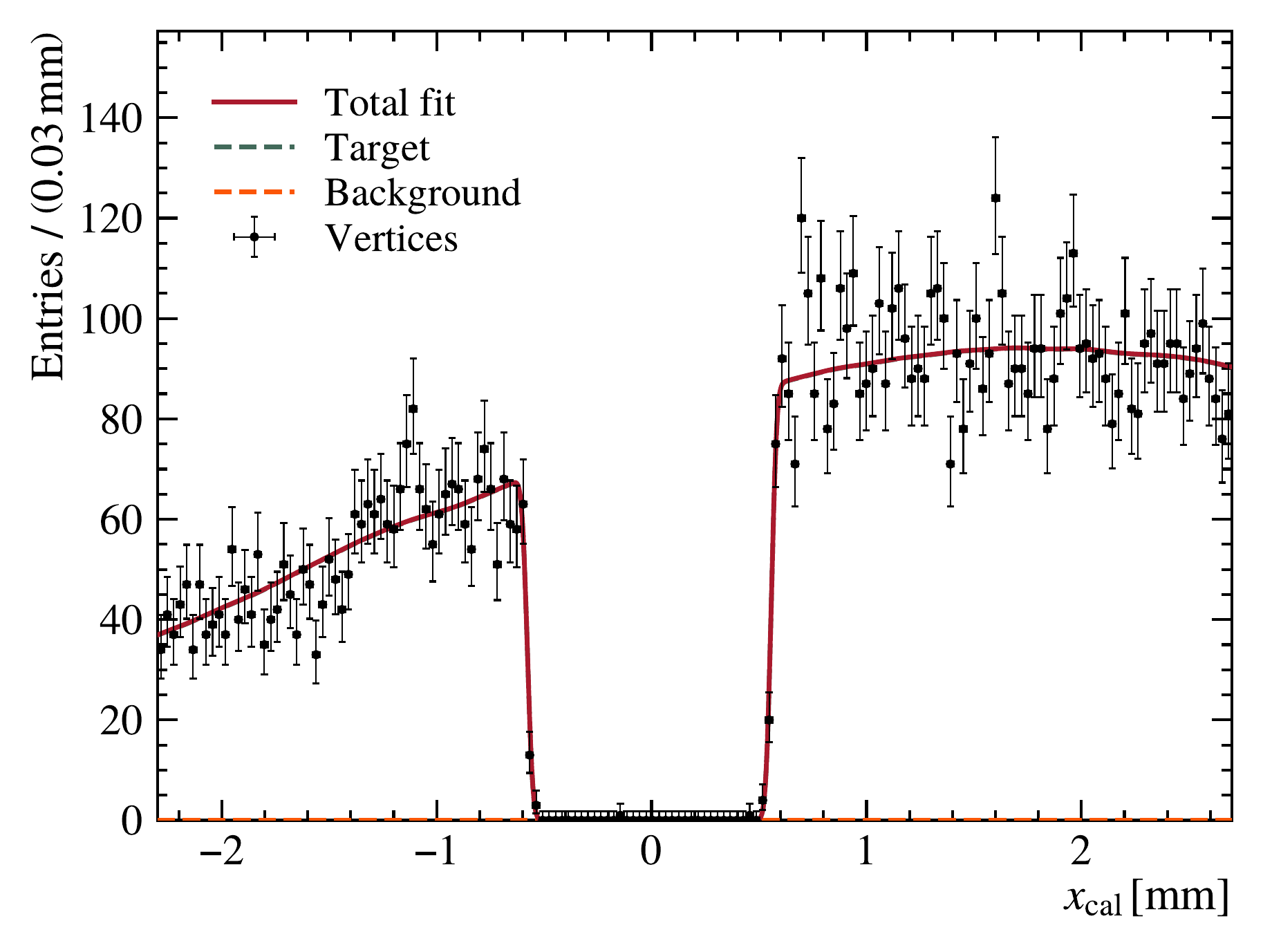}
        \caption{}\label{fig:spatial_resolution:transverse:fit:x5tracks}
    \end{subfigure}
    \caption{Fits to the $x_{\text{cal}}$ distribution of vertices with \subref{fig:spatial_resolution:transverse:fit:x2tracks} two and \subref{fig:spatial_resolution:transverse:fit:x5tracks} five tracks.
    }\label{fig:spatial_resolution:transverse:fit}
\end{figure}

Example fits to the $x_{\text{cal}}$ distribution for vertices with two and five tracks are shown in \cref{fig:spatial_resolution:transverse:fit}.
An equivalent model is used for the $y$-oriented slit, fitting $y_{\text{cal}}$ with resolution parameter $\sigma_y$, with the fits giving similar results.

\begin{figure}
    \centering
    \begin{subfigure}{0.5\textwidth}
        \centering
        \includegraphics[width=0.9\textwidth]{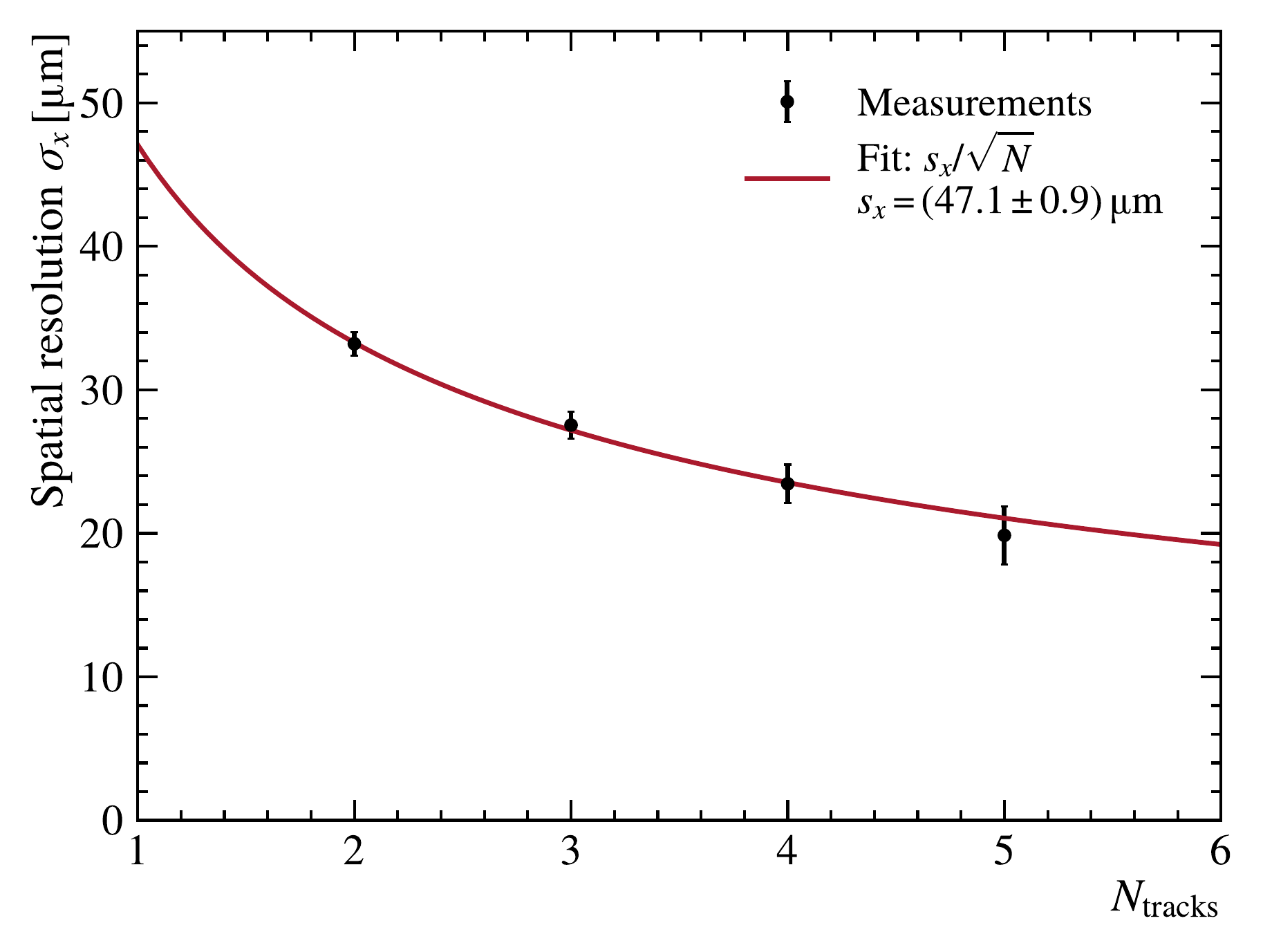}
        \caption{}\label{fig:spatial_resolution:transverse:resolution:x}
    \end{subfigure}%
    \hfill%
    \begin{subfigure}{0.5\textwidth}
        \centering
        \includegraphics[width=0.9\textwidth]{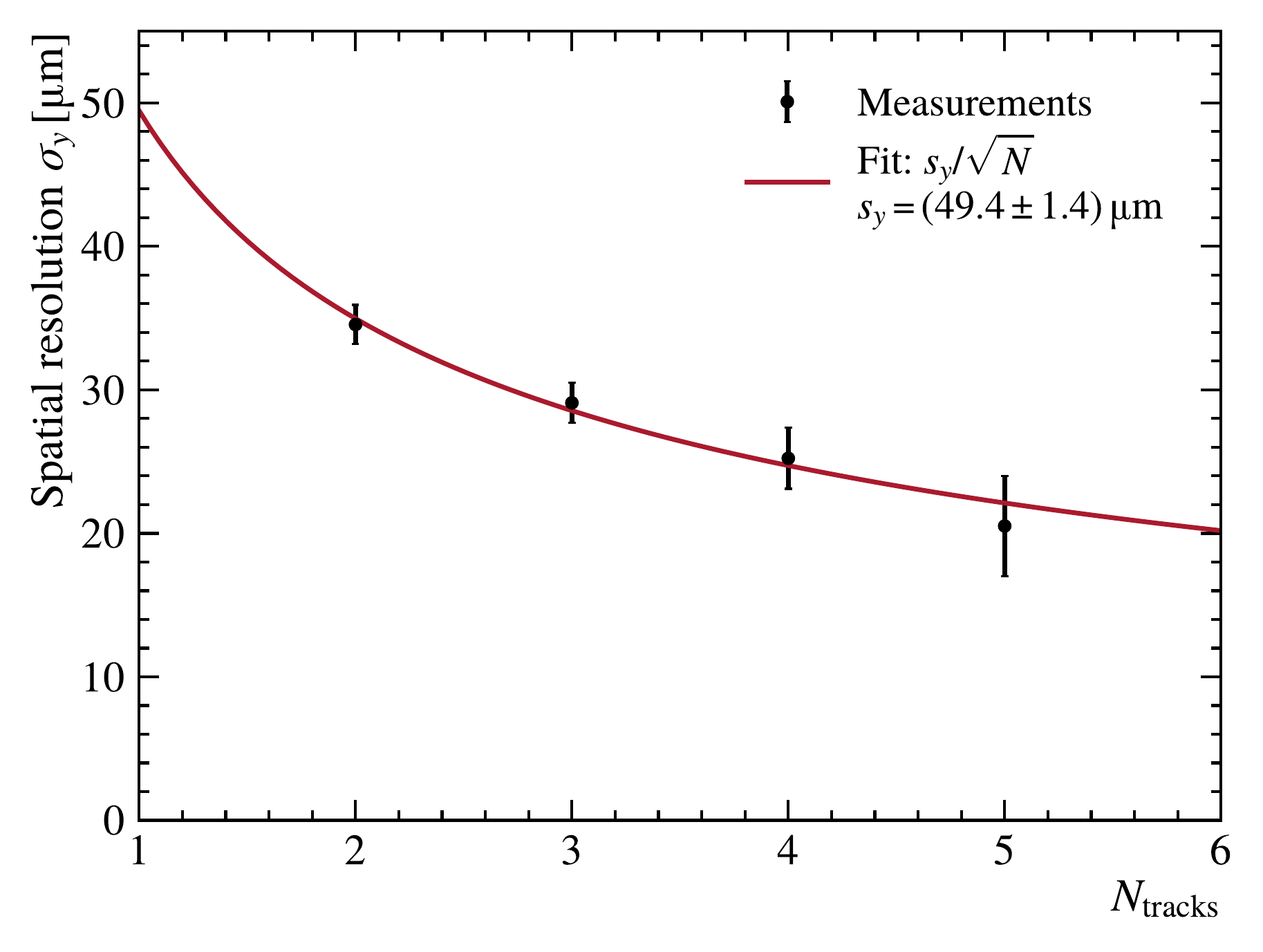}
        \caption{}\label{fig:spatial_resolution:transverse:resolution:y}
    \end{subfigure}
    \caption{Transverse resolutions \subref{fig:spatial_resolution:transverse:resolution:x} $\sigma_x$ and \subref{fig:spatial_resolution:transverse:resolution:y} $\sigma_y$ as a function of $N_{\text{tracks}}$, with best fits to $\sigma_{x,y} = s_{x,y}/\sqrt{N_{\text{tracks}}}$ overlaid.
    }\label{fig:spatial_resolution:transverse:resolution}
\end{figure}

The fitted transverse resolutions $\sigma_x$ and $\sigma_y$ are shown as functions of $N_{\text{tracks}}$ in \cref{fig:spatial_resolution:transverse:resolution}.
Both reach approximately \qty{20}{\micro\meter} for five-track vertices.
In contrast to the longitudinal resolution, each track contributes an independent measurement of the transverse vertex position, and the resolution is therefore well described by
\begin{equation}
    \sigma_{x,y}(N_{\text{tracks}}) = \frac{s_{x,y}}{\sqrt{N_{\text{tracks}}}},
\end{equation}
yielding best-fit values of $s_x = \qty{47.1\pm0.9}{\micro\meter}$ and $s_y = \qty{49.5\pm1.4}{\micro\meter}$, with $\chisqndf =\num{0.17}$ for both fits.
The inferred single-track transverse resolutions are somewhat larger than the expected pointing resolution of approximately \qty{35}{\micro\meter} at the target position shown in \cref{fig:setup:simulation:track_resolution}.
This discrepancy is likely caused by the same effects that produce the non-Gaussian tails in the longitudinal vertex distribution.
Both the uncertainty in the per-track momentum estimates and the unmodelled material in the corners of the telescope planes degrade the track pointing resolution.

\section{Temporal resolution}\label{sec:temporal_resolution}
The time resolution of the reconstructed vertices is studied using the same tungsten target configuration as for the longitudinal spatial resolution measurement.
To increase the size of the data sample, the study includes vertices reconstructed from interactions in all material upstream of the telescope, as shown in \cref{fig:reconstruction:vertex_distribution}, rather than only those in the tungsten layers.
Throughout this section, time resolutions are defined as the RMS of the central \qty{99}{\percent} of the relevant residual distribution with respect to the reference time.
The \qty{12}{\pico\second} contribution from the MCP time reference is subtracted in quadrature.

\subsection{Single-plane time resolution}
\begin{figure}
    \centering
    \begin{subfigure}{0.5\textwidth}
        \centering
        \includegraphics[width=0.9\textwidth]{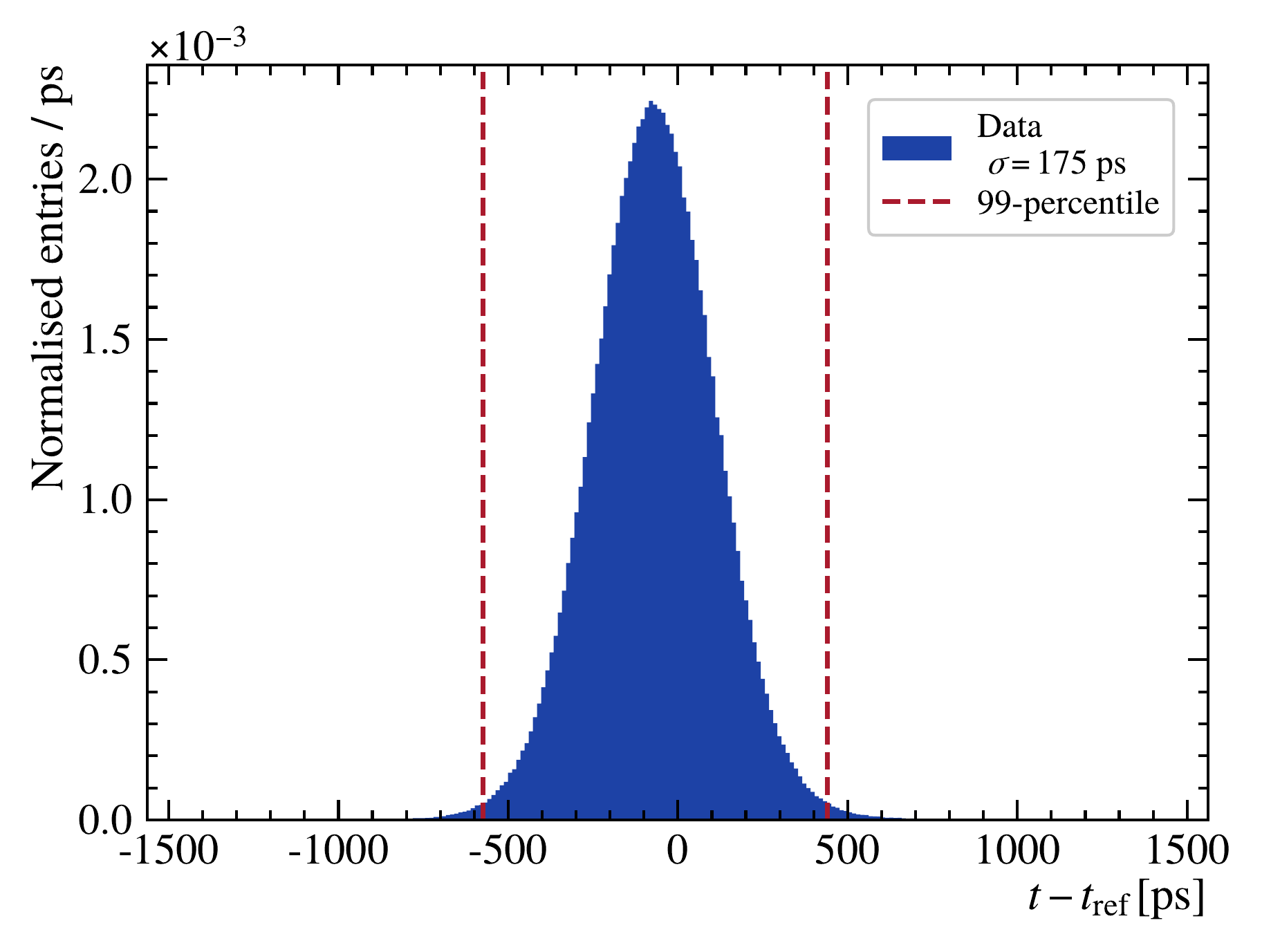}
        \caption{}\label{fig:temporal_resolution:time_residuals:timing}
    \end{subfigure}%
    \hfill%
    \begin{subfigure}{0.5\textwidth}
        \centering
        \includegraphics[width=0.9\textwidth]{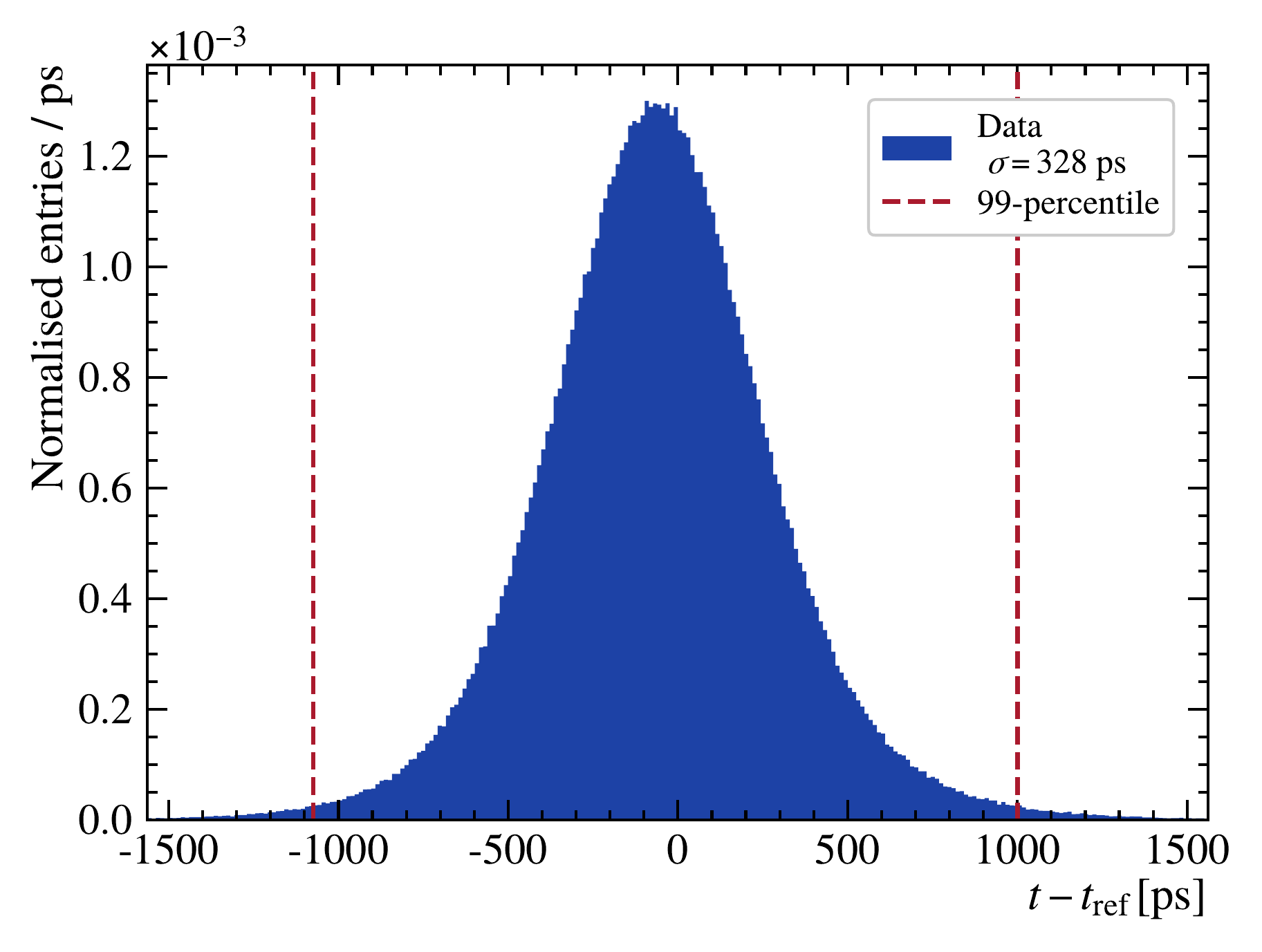}
        \caption{}\label{fig:temporal_resolution:time_residuals:spatial}
    \end{subfigure}%
    \caption{Time residuals with respect to the MCP time reference for a representative \subref{fig:temporal_resolution:time_residuals:timing} timing and \subref{fig:temporal_resolution:time_residuals:spatial} spatial plane.
    }\label{fig:temporal_resolution:time_residuals}
\end{figure}

The per-plane time residuals for vertex tracks with respect to the MCP time reference are shown in \cref{fig:temporal_resolution:time_residuals} for representative timing and spatial planes.
The extracted resolutions are slightly worse than those obtained for isolated beam tracks.
The timing-plane resolutions range from \qty{175}{\pico\second} to \qty{195}{\pico\second}, compared with \qtyrange{160}{180}{\pico\second} for beam tracks, while the spatial-plane resolutions are approximately \qty{330}{\pico\second}, compared with \qty{275}{\pico\second}.
This degradation is largely attributed to the broad angular distribution of the secondary particles, which increases inter-pixel charge sharing and introduces additional timewalk effects that are not fully corrected by the neural network--based calibration.
Although the track slopes are included as inputs to the MLP, the training sample is dominated by small-angle isolated beam tracks, so the corrections are less accurate for the large-angle vertex tracks.
The pointing resolution at each plane is also degraded for vertex tracks, which further reduces the accuracy of the timewalk corrections via the intrapixel intercept input.

\subsection{Vertex time resolution}
\begin{figure}
    \centering
    \begin{subfigure}{0.5\textwidth}
        \centering
        \includegraphics[width=0.8\textwidth]{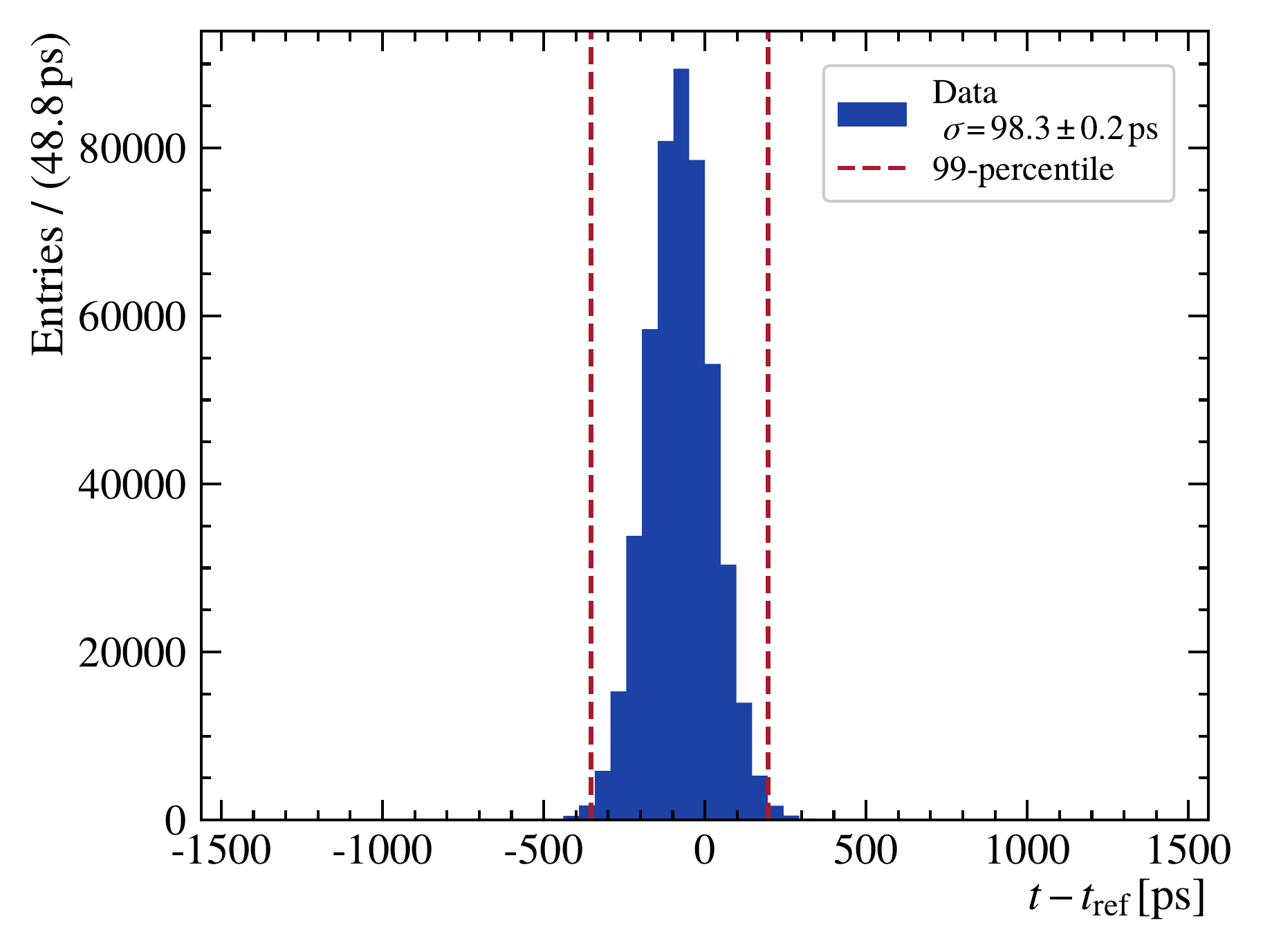}
        \caption{}\label{fig:temporal_resolution:vertex_time_residuals:n4}
    \end{subfigure}%
    \hfill%
    \begin{subfigure}{0.5\textwidth}
        \centering
        \includegraphics[width=0.8\textwidth]{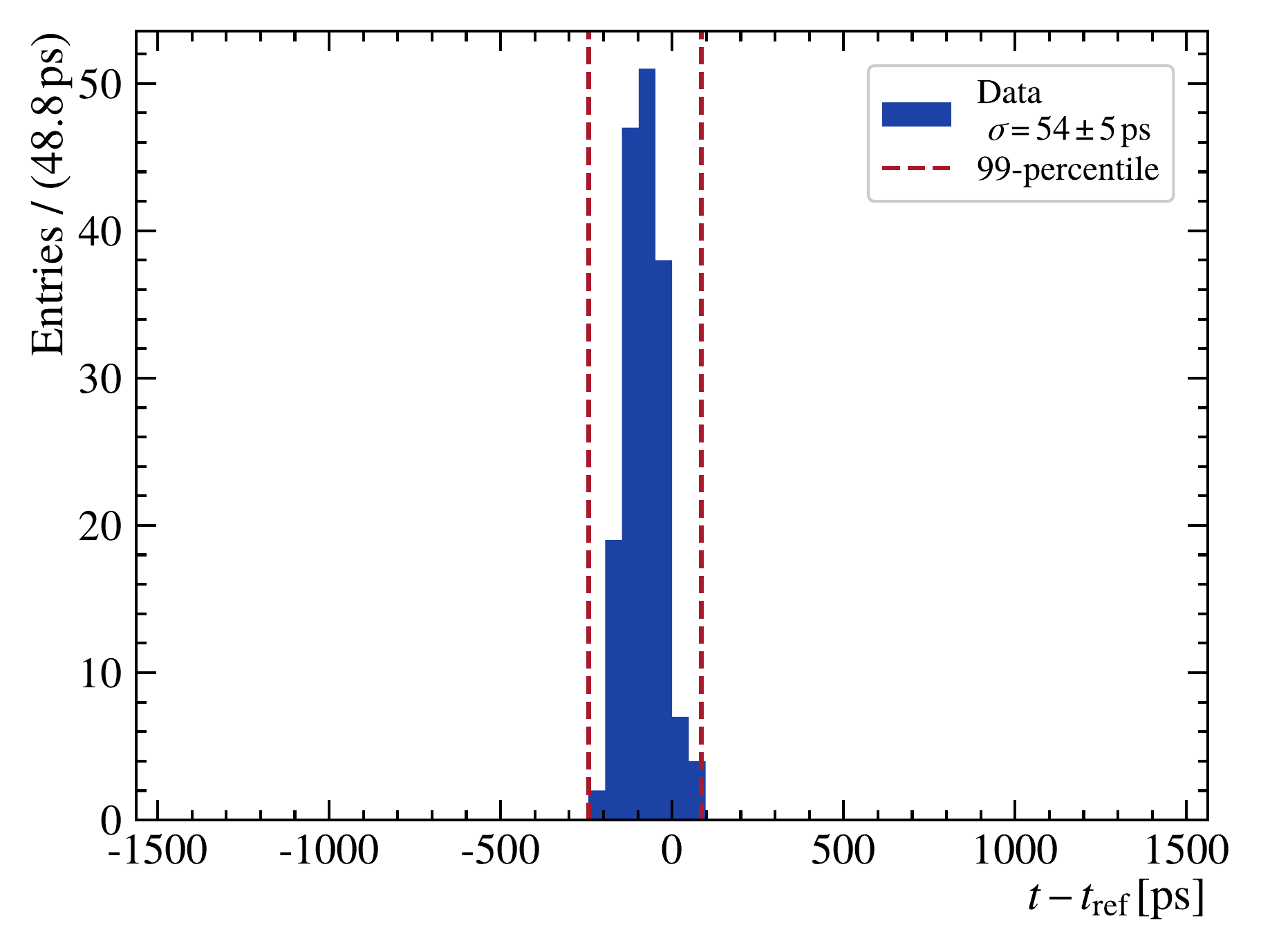}
        \caption{}\label{fig:temporal_resolution:vertex_time_residuals:n16}
    \end{subfigure}%
    \caption{Vertex time residuals with respect to the MCP time reference for vertices with \subref{fig:temporal_resolution:vertex_time_residuals:n4} four and \subref{fig:temporal_resolution:vertex_time_residuals:n16} sixteen timing clusters.
    }\label{fig:temporal_resolution:vertex_time_residuals}
\end{figure}

The vertex time resolution is studied as a function of the number of timing-plane clusters, $N_{\text{timing}}$, contributing to the fitted vertex time.
This parametrisation is preferred to the vertex track multiplicity because each track may be reconstructed on a different subset of planes.
A track confined to the front arm contributes fewer timing measurements than one traversing the full telescope, so vertices with the same track multiplicity may contain substantially different amounts of timing information.
Contributions from spatial planes are suppressed by artificially inflating their time resolutions in the track and vertex fits to \qty{10}{\nano\second}, rendering them effectively weightless in the vertex time fit.
For each value of $N_{\text{timing}}$ between one and 16, the resolution is extracted from the vertex time residuals with respect to the MCP time reference.
Example residual distributions for vertices with four and sixteen timing clusters are shown in \cref{fig:temporal_resolution:vertex_time_residuals}.

\begin{figure}
    \centering
    \includegraphics[width=0.6\textwidth]{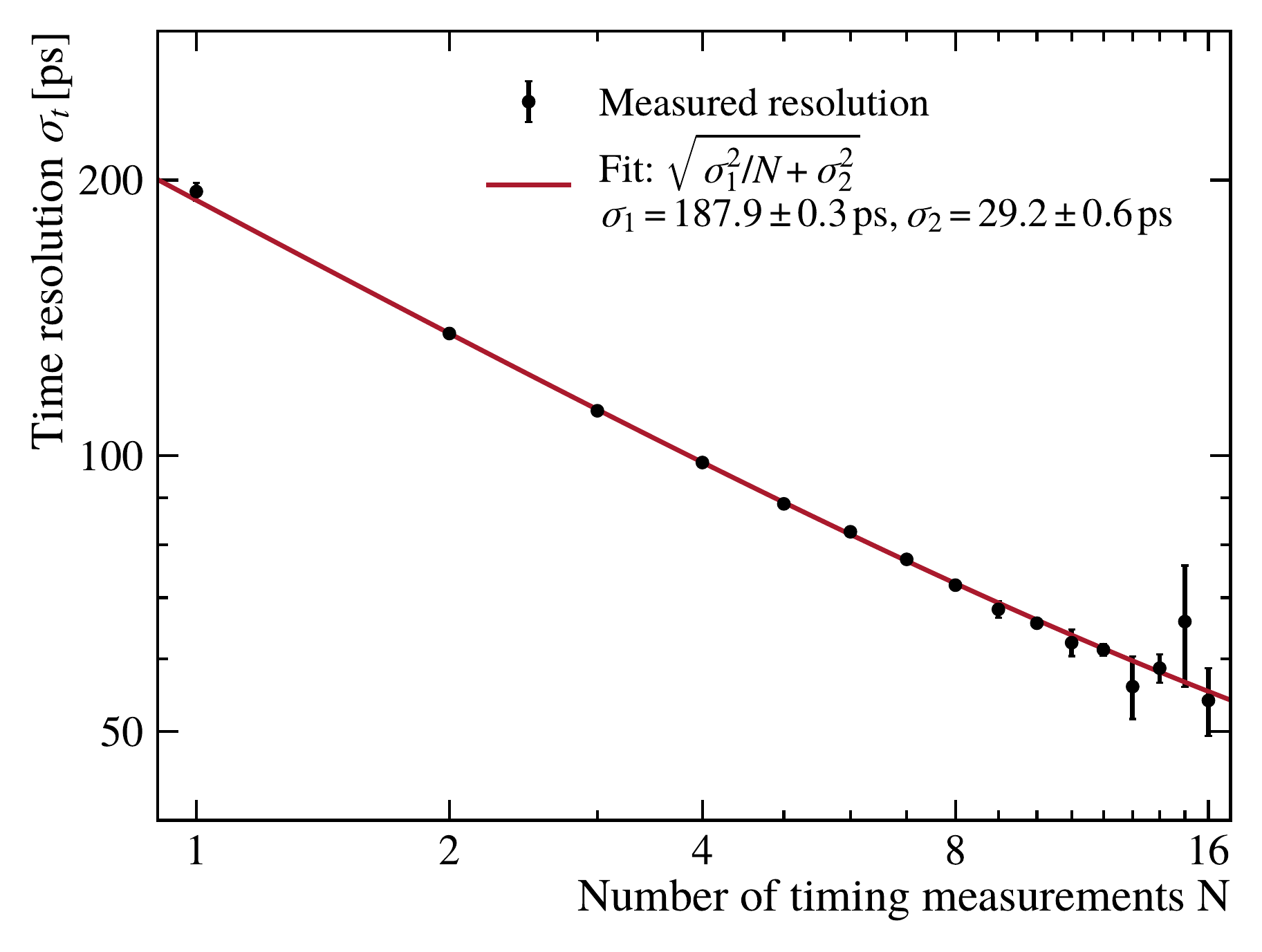}
    \caption{Vertex time resolution as a function of the number of timing-plane clusters $N_{\text{timing}}$, with the best fit to $\sigma_t = \sqrt{\sigma_1^2/N_{\text{timing}} + \sigma_2^2}$ overlaid.
    The uncertainties shown have been scaled by the Birge ratio $\sqrt{\chisqndf} \approx \num{1.47}$.
    }\label{fig:temporal_resolution:vertex_time_resolution}
\end{figure}

The resulting vertex time resolution is shown as a function of $N_{\text{timing}}$ in \cref{fig:temporal_resolution:vertex_time_resolution}.
It improves with the number of timing-plane clusters, reaching \qty{54\pm5}{\pico\second} for vertices with 16 timing clusters.
The resolution is well described by a function of the form
\begin{equation}
    \sigma_t(N_{\text{timing}}) = \sqrt{\frac{\sigma_1^2}{N_{\text{timing}}} + \sigma_2^2},
\end{equation}
where $\sigma_1$ is the effective single-plane time resolution and $\sigma_2$ is an irreducible contribution that does not improve with additional timing measurements.
The fit yields $\sigma_1 = \qty{187.9\pm0.3}{\pico\second}$ and $\sigma_2 = \qty{29.2\pm0.6}{\pico\second}$,  with $\chisqndf = \num{2.2}$.
The uncertainties on the vertex time resolutions shown in \cref{fig:spatial_resolution:longitudinal:resolution}, and on $\sigma_1$ and $\sigma_2$ are therefore scaled by the  Birge ratio $\sqrt{\chisqndf} \approx \num{1.47}$.
The single-plane resolution, $\sigma_1$, is consistent with the one extracted from the time residuals of the individual planes, showing that timestamps from multiple detector planes and tracks combine according to the expected statistical scaling.
The irreducible component, $\sigma_2$, contains a contribution of less than \qty{15}{\pico\second} from the jitter in the clock distribution~\cite{Timepix4Telescope:2025}.
Additionally, it receives a contribution up to \qty{25}{\pico\second} due to a positional dependence of timestamp of the time reference system.
The time at which the Cherenkov photons reach the MCP depends on the horizontal impact position on the quartz fingers, due to the \qty{45}{\degree} angle and the time of flight of the photons inside the quartz.
Over the width of a Timepix4 ASIC this results in a variation of the average timestamp of approximately \qty{25}{\pico\second}, which does not appear in the time difference of the two MCP signals.
This variation will be reduced since the vertices combine multiple tracks, but is most likely still the main contribution to $\sigma_2$, besides the clock jitter.

\section{Conclusion}\label{sec:conclusion}
The first demonstration of full 4D vertex reconstruction at the \qty{50}{\pico\second} scale using a silicon pixel detector system is presented.
Hadronic interactions of a \qty[per-mode=symbol]{180}{\giga\electronvolt\per\clight} mixed-hadron beam were reconstructed with the \timepixfour{} beam telescope.
Cluster time measurements from multiple detector planes were combined in Kalman filter--based track fits, and the fitted track states were subsequently combined into vertices.
In all fits, time was included directly in both the track and vertex states and fitted simultaneously with the spatial parameters rather than assigned after spatial reconstruction.

The vertex time resolution was evaluated as a function of the number of timing-plane clusters contributing to the fitted vertex time, after subtracting the known \qty{12}{\pico\second} contribution from the time reference system.
It improves according to the expected statistical scaling, reaching \qty{54\pm5}{\pico\second} for vertices with \num{16} timing clusters.
The fitted effective single-plane resolution of \qty{187.9\pm0.3}{\pico\second} is consistent with the measured detector performance, showing that the timing measurements combine as expected.
An irreducible contribution of \qty{29.2\pm0.6}{\pico\second} remains in the fitted resolution model, including large contributions from the clock distribution and from a positional dependence of the time reference system rather than correlations among the timing measurements themselves.
Additionally, the spatial resolution of the reconstructed vertices exhibits the expected dependence on the track multiplicity, reaching approximately \qty{20}{\micro\meter} transversely for five-track vertices and \qty{660}{\micro\meter} longitudinally for six-track vertices.

Although no pile-up was present in these measurements, and they do not yet reach the timing performance of \qty{20}{\pico\second} envisaged for the \lhcb{} \upgradetwo, they establish that timestamps form multiple silicon pixel detector planes and tracks can be combined to achieve vertex time resolutions at the \qty{50}{\pico\second} scale without an unforeseen resolution floor above it.
This represents a significant experimental step towards exploiting precision timing for vertex association in future high-pile-up detectors.

\section*{Acknowledgements}
We thank our colleagues in the \cern accelerator departments for the excellent performance of the beam in the SPS North Area.
We thank Miranda van Stenis and  Kacper Jama for their help in the preparation of the time reference's mechanics, and Thierry Gys for his guidance and support during the early stages of operation of the MCP-PMTs.
We acknowledge the ASIC designers of \timepixfour and PicoTDC, and the designers of the SPIDR4 readout system.
We gratefully acknowledge the computing resources and batch services provided by \cern.
We also thank the open-source software communities that supported this work.
We acknowledge the \cern EP R\&D program's contribution to the construction of the \timepixfour beam telescope. 
We also gratefully acknowledge the support from the Netherlands Organisation for Scientific Research (NWO),
and the Science and Technology Facilities Council (UK, grant number ST/V003151/1).

\clearpage
\addcontentsline{toc}{section}
{References}
\setboolean{inbibliography}{true}

\bibliographystyle{elsarticle-num-names}
\bibliography{main}
\end{document}